\documentclass{aa}

\usepackage[varg]{txfonts}

\usepackage{savesym}
\savesymbol{iint}
\savesymbol{iiint}
\savesymbol{idotsint}
\savesymbol{iiiint}

\usepackage{amsmath, amssymb}
\usepackage[dvips]{graphicx}
\usepackage{epsfig}
\usepackage{natbib}

\newcommand{\Blight}{B_{\rm L}}
\newcommand{\rlight}{r_{\rm L}}

\graphicspath{
{/home/petri/simulation/pulsar/vent/emission/}
}

\begin{document}

\title{Effect of geodetic precession on the evolution of pulsar high-energy pulse profiles as derived with the striped-wind model.}
\titlerunning{Geodetic precession in the pulsar striped-wind}

\author{J. P\'etri\thanks{E-mail: jerome.petri@astro.unistra.fr}}

\institute{Observatoire Astronomique de Strasbourg, Universit\'e de Strasbourg, CNRS, UMR 7550, 11 rue de l'Universit\'e, 67000 Strasbourg, France.}

\date{Received / Accepted }

\abstract{Geodetic precession has been observed directly in the double-pulsar system PSR~J0737-3039. Its rate has even been measured and agrees with predictions of general relativity. Very recently, the double pulsar has been detected in X-rays and gamma-rays. This fuels the hope observing geodetic precession in the high-energy pulse profile of this system. Unfortunately, the geometric configuration of the binary renders any detection of such an effect unlikely. Nevertheless, this precession is probably present in other relativistic binaries or double neutron star systems containing at least one X-ray or gamma-ray pulsar.}{We compute the variation of the high-energy pulse profile expected from this geodetic motion according to the striped-wind model. We compare our results with two-pole caustic and outer gap emission patterns.}{For a sufficient misalignment between the orbital angular momentum and the spin angular momentum, a significant change in the pulse profile as a result of geodetic precession is expected in the X-ray and gamma-ray energy band.}{The essential features of the striped wind are indicated in several plots showing the evolution of the maximum of the pulsed intensity, the separation of both peaks, if present, and the variation in the width of each peak. We highlight the main differences with other competing high-energy models.}{We make some predictions about possible future detection of high-energy emission from double neutron star systems with the highest spin precession rate. Such observations will definitely favour some pulsed high-energy emission scenarios.}


\maketitle

\section{INTRODUCTION}

The discovery of more than 150~gamma-ray pulsars by Fermi/LAT (Large Area Telescope) opens up a new window to probe the energetics of such stars. Indeed, contrary to other wavelengths, many gamma-ray pulsars have a considerable gamma-ray luminosity efficiency and sometimes close to the spin-down luminosity itself \citep{2013ApJS..208...17A}. Therefore, high-energy observations can help perform a calorimetric investigation on the kinetic energy losses of pulsars. This gives us valuable information about the radiation mechanisms and their location within the magnetosphere or wind, especially the geometry as reported by \citet{2009ApJ...707..800V} and \citet{2010ApJ...714..810R}. The shape of the pulse profiles in gamma-rays as well as in the other wavebands depends on the particular geometry, that is, on the inclination angle of the line of sight with respect to the spin axis and the magnetic obliquity of the neutron star. These quantities are usually unknown and must be fitted for any model trying to reproduce the observations. Several interpretations of the Fermi/LAT pulse profiles have been reported but they cannot be distinguished because of the degeneracy between models. 

Our idea in this paper is to face this problem from a different side and try to lift this degeneracy by looking for a possible secular evolution of the high-energy pulse profiles caused by the changing viewing geometry resulting from geodetic precession. Whereas several models of high-energy emission can satisfactorily reproduce pulse profiles for {\it a fixed geometry} including the inclination of the line of sight and the magnetic obliquity, they will differ by the imprint on the pulse profiles from an evolving inclination angle of the line of sight. Indeed, the caustic nature of the intensity maps extracted from magnetospheric models used by \citet{2010ApJ...715.1282B} differ significantly from the S-shape intensity maps obtained by the striped pulsar wind pulse profile that are reported by  \citet{2011MNRAS.412.1870P}. One possible means to lift this degeneracy between competing models is to look for secular variation in the geometric configuration. Such effects could come from geodetic precession. It is well known that general relativity predicts a spin-orbit coupling that could be detectable via spin precession \citep{1975PhRvD..12..329B, 1975A&A....44..417B, 1988A&A...202..109B}. This has indeed been observed in several binary pulsars in the evolution of their radio pulses. \cite{1989ApJ...347.1030W} for instance have observed geodetic precession in the historical pulsar PSR~B1913+16. The period of precession was predicted to be 180~years and allowed surveying a significant fraction of the precession cycle in a human lifetime. In this particular case, six years were sufficient to see a change in the pulse profile. Polarization data also constrain the precessing motion of the pulsar, as explained by \cite{1990ApJ...349..546C}. After twenty years of observations, \cite{2002ApJ...576..942W} were even able to map the two-dimensional geometry of the radio-emitting beam. The double pulsar PSR~J0737$-$3039A/B shows the fastest precession rate with periods of 71 years for pulsar~A and 74 years for pulsar B \citep{2008Sci...321..104B}. Recently, pulsar~A has been detected in gamma-rays by Fermi/LAT \citep{2013ApJ...768..169G}, but its rotation axis is not sufficiently inclined with respect to the orbital angular momentum to detect a noticeable precession \citep{2013ApJ...767...85F}. This was already claimed by \citet{2005ApJ...621L..49M}, who did not find any relevant change of pulse profile in pulsar A over several years. Nevertheless, \citet{2005ApJ...634.1223L} showed how the eclipses of pulsar~A can help in constraining its geometry thanks to a simple magnetospheric model assuming a dipolar field. Even after several years, the statistics will not reach a satisfactory level to enable measuring the variation in the pulse profiles. However, the double pulsar is not the only pulsar binary that displays spin precession. For instance, several other pulsars have since been observed to show geodetic precession, such as the relativistic binary pulsars PSR~B1534+12, as reported by \cite{2003ApJ...589..495K}, and PSR~J1141-6545 described by \cite{2005ApJ...624..906H, 2010ApJ...710.1694M}. From a more fundamental point of view, determining the precession rate offers another test of gravity theories, as shown by \cite{2008Sci...321..104B}. Measurements of the misalignment between the pulsar spin axis and the orbital momentum axis give some insight into the evolution of the binary pulsar. It is probably related to an asymmetric kick during the supernova explosion \citep{2013ApJ...767...85F, 2004ApJ...616..414W, 2005ApJ...619.1036T}.

Here we discuss in detail the effect of geodetic precession on the secular evolution of the high-energy pulse profiles in the striped-wind interpretation. In Sect.~\ref{sec:Modele}, we outline this striped-wind model, using the latest improvements, to compute the light curves. Then in Sect.~\ref{sec:Results}, we show the results of geodetic precession on the pulse profile, its separation between the two peaks (if present), and the variation in the width of the pulses. Discussion and possible near (or not so near) future detections are discussed in Sect.~\ref{sec:Discussion} for some candidate binary neutron stars. We also report the main characteristics of the two-pole caustic and outer-gap model for comparison with the striped-wind model. Finally in Sect.~\ref{sec:Conclusion} we conclude our work and offer ideas for possible future investigations of this phenomenon in pulsars.

\section{STRIPED-WIND GEOMETRY AND GEODETIC PRECESSION}
\label{sec:Modele}

In this section, we briefly summarize the striped-wind model that was used to compute the light curves. The magnetic field structure follows the same geometry as the one used to derive the pulsed synchrotron polarization features of the wind \citep{2013MNRAS.434.2636P}. We then give the exact expression for the line-of-sight evolution caused by geodetic precession.

\subsection{Magnetic field structure and particle distribution function}

The exact analytical solution of the electromagnetic field in a finite-thickness striped pulsar wind expanding radially outwards at a constant speed~$V=\beta_{\rm v}\,c$ slightly lower than the speed of light denoted by~$c$ was given in \cite{2013MNRAS.434.2636P}. For completeness, we recall this structure briefly. In spherical polar coordinates $(r,\vartheta,\varphi)$ centred on the star and with the $z$-axis along the rotation axis, the explicit expressions for the electromagnetic field components in the rest frame of the star are (the same as the observer frame)
\begin{subequations}
  \label{eq:Champ_EM_Strie}
\begin{align}
  B_r & = \beta_{\rm v}^2 \, \Blight \, \frac{\rlight^2}{r^2} \, \tanh (\Psi_s/\Delta) \\
  B_\vartheta & = 0 \\
  B_\varphi & = - \beta_{\rm v} \, \Blight \, \frac{\rlight}{r} \, \sin \vartheta \, \tanh (\Psi_s/\Delta) \\
  E_r & = 0 \\
  E_\vartheta & = - \beta_{\rm v}^2 \, c \, \Blight \, \frac{\rlight}{r} \, \sin \vartheta \, \tanh (\Psi_s/\Delta) \\
  E_\varphi & = 0 .
\end{align}
\end{subequations}
Here, $\rlight=c/\Omega$ is the radius of the light-cylinder, $\Omega$ is the angular velocity of the pulsar, $\Blight$ is a fiducial magnetic field strength in the vicinity of the light-cylinder, and $\Delta$ represents a parameter quantifying the length scale of the current sheet thickness. This current sheet is located in regions where the function
\begin{equation}
  \label{eq:PSI_S}
  \Psi_s = \cos \vartheta \, \cos \chi + \sin \vartheta \, \sin \chi \, \cos\left[\varphi - \Omega \, ( t - \frac{r}{\beta_{\rm v}\,c} ) \right]
\end{equation}
is nearly zero, $\chi$ is the obliquity, that is, the angle between the magnetic and rotation axes. These analytical closed expressions for the striped wind with a current sheet of finite extent satisfy the homogeneous Maxwell equations. Note also that the magnetic structure does not possess the property $B_r = B_\varphi$ in the equatorial plane of the light-cylinder. The ratio of their magnitude at that point is equal to~$\beta_{\rm v}=V/c$. Nevertheless, this solution is physically satisfactory because it satisfies the constrain $E<c\,B$ everywhere in space, as shown in \cite{2013MNRAS.434.2636P}. In our picture, the current sheet is filled with a dense and hot weakly magnetized plasma and is surrounded by a cold and diluted strongly magnetized plasma. We mainly expect synchrotron radiation from particles trapped in the hot current sheet producing gamma-ray photons in the Fermi/LAT range \citep{2012MNRAS.424.2023P}. We recall that the resulting pulsed emission is due to the combined effect of the spiral structure of the current sheet; which  expands at speeds close to the light velocity and relativistic beaming \citep{2002A&A...388L..29K}. The observed pulse profile reflects the geometry of the current sheet. The pulse width is uniquely related to the thickness of the current sheet, see \cite{2011MNRAS.412.1870P}.
The magnetic field structure presented in equation~(\ref{eq:Champ_EM_Strie}) is reminiscent of the asymptotic split monopole solution given for two half monopolar magnetic fields connected at the stellar surface. At first sight, it seems that this solution is only applicable far from the light-cylinder. However, according to recent 3D numerical simulations of the extended pulsar magnetosphere by \cite{2012MNRAS.420.2793K} and from the orthogonal rotator by \cite{2012MNRAS.424..605P}, this split monopole geometry forms close to the light-cylinder. The finite thickness striped wind of equation~(\ref{eq:Champ_EM_Strie}) is therefore a good approximation for $r\gtrsim\rlight$. Nevertheless, in our computations, we used $r\ge10\,\rlight$ where the monopolar field is expected to accurately represent the true field irrespective of the transition between the closed magnetosphere and the wind around the light-cylinder.

Since the first computation of the phase-resolved polarization properties of the striped wind reported by \cite{2005ApJ...627L..37P} for synchrotron emission, the model has evolved towards inverse Compton emission to explain for instance Geminga pulsar phase-resolved spectra \citep{2009A&A...503...13P} and the variety of phase plots \citep{2011MNRAS.412.1870P}. Recently, \cite{2012MNRAS.424.2023P} invoked synchrotron emission to account for pulsar gamma-ray luminosities as observed by the Fermi/LAT instrument. In the present work, the emission model is based on the synchrotron radiation mechanism emanating from the finite-thickness striped wind.

\subsection{Secular evolution caused by geodetic precession}

Because all the details of the geometry, such as the obliquity of the pulsar~$\chi$ and inclination of the line of sight~$\zeta$, magnetic field structure, and particle distribution functions, have already been reported in \cite{2013MNRAS.434.2636P}, we refer to this paper for a precise definition of all physical quantities. We do not reproduce them here to avoid redundancy. Now the novelty comes from assuming that the pulsar belongs to a binary system and that its orbital motion and spin suffer significantly from general-relativistic effects such as geodetic precession. Therefore we assume that the orbital plane of the system is inclined with an angle~$i$ with respect to the line of sight. The orbital angular momentum is denoted by~$\textbf{\textit{L}}$, the spin angular momentum of the pulsar by~$\mathbf \Omega$. General relativity predicts that the neutron star will precess about the total angular momentum of the system, but because the pulsar angular spin momentum is much smaller than the orbital angular momentum, to a very good accuracy, the neutron star will only precess around~$\textbf{\textit{L}}$. If $\delta$ is the angle formed by the two vectors $\textbf{\textit{L}}$ and $\mathbf{\Omega}$, geodetic precession induces a secular change in the inclination of the observer line of sight such that
\begin{equation}
\label{eq:zetaPrecession}
 \cos \zeta = \cos \delta \, \cos i + \sin \delta \, \sin i \, \cos \varphi.
\end{equation}
For the definition of the different angles, see for instance \citet{1992PhRvD..45.1840D}. Note the symmetric role played by orbital inclination~$i$ and misalignment~$\delta$. Their interchange does not affect the variation in the inclination of the line of sight~$\zeta$. The precession phase~$\varphi$ is related to geodetic precession by $\varphi = \Omega_{\rm p}\,(t-t_0)$, where $t_0$ is a reference time for which the orbital angular momentum, the pulsar spin axis, and the observer line of sight are in a same plane. The rate of precession according to general relativity is given by
\begin{equation}
\label{eq:Precession}
 \Omega_{\rm p} =  \left( \frac{2\,\pi}{P_b}\right) ^{5/3} \, T_\odot^{2/3} \, \frac{1}{1-e^2} \, \frac{m_c\,(4\,m_p+3\,m_c)}{2\,(m_p+m_c)^{4/3}},
\end{equation}
where we introduced an unit of time such that
\begin{equation}
 T_\odot = \frac{G\,M_\odot}{c^3} \approx4.925~\mu s.
\end{equation}
$G$ represents the gravitational constant, $c$ the speed of light, $M_\odot$ the mass of the Sun, $P_b$ the orbital period of the pulsar, $e$ the eccentricity of the orbit, $m_p$ the mass of the pulsar, and $m_c$ the mass of its companion.
Note that the definition of the angles in equation~(\ref{eq:zetaPrecession}) can differ from author to author, especially the inclination of the orbital plane with respect to the line of sight~$i$. Our definitions are consistent with the convention adopted by \cite{1998ApJ...509..856K}.

\section{RESULTS}
\label{sec:Results}

We now show a sample of pulse-profile evolutions with different geodetic precession assumed. Time is normalized to one period of precession and can be easily converted into physical units through equation~(\ref{eq:Precession}). We discuss the main features of the pulse profiles, such as the separation of the double pulse profile, if two pulses are present, the variation in pulse width, and the highest intensity. Table~\ref{tab:Parametres} summarizes the four cases investigated in depth.
\begin{table}
 \centering
 \begin{tabular}{cccc}
\hline
  Case & $\chi$ & $\delta$ & $i$ \\
\hline
\hline
  1 & $30^o$ & $30^o$ & $60^o$ \\
  2 & $60^o$ & $30^o$ & $60^o$ \\
  3 & $60^o$ & $30^o$ & $90^o$ \\
  4 & $90^o$ & $60^o$ & $30^o$ \\
\hline
 \end{tabular}
 \caption{The four cases with their corresponding angles~$(\chi,\delta,i)$ as envisaged in this work.}
 \label{tab:Parametres}
\end{table}

\subsection{Theoretical expectations}

\begin{figure}
 \centering
  \includegraphics[width=0.45\textwidth]{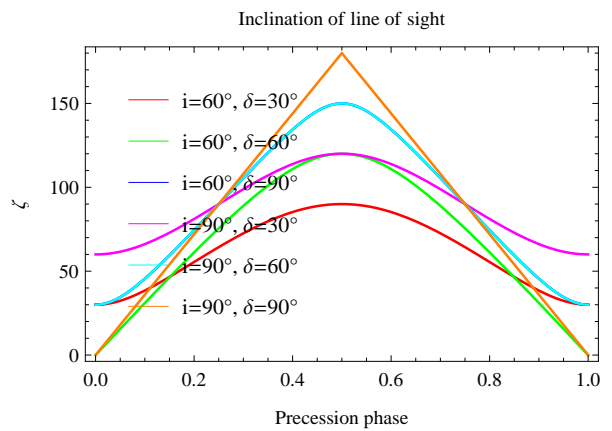} \\
 \caption{Evolution of the inclination of the line of sight~$\zeta$ with respect to the precession phase~$\varphi$ for different misalignment angles~$\delta$ and orbital plane inclinations~$i$. The geometrical parameters are shown in the legend, they are $\delta=\{30^o, 60^o, 90^o\}$, $i=\{60^o, 90^o\}$. Only $(i,\delta)=(90^o,90^o)$ gives a full $180^o$ switch in $\zeta$. $(i,\delta)=(60^o,90^o)$ is equal to $(i,\delta)=(90^o,60^o)$ because of symmetry and therefore cannot be distinguished in the plot.}
 \label{fig:Zeta}
\end{figure}

\begin{figure}
 \centering
  \includegraphics[width=0.45\textwidth]{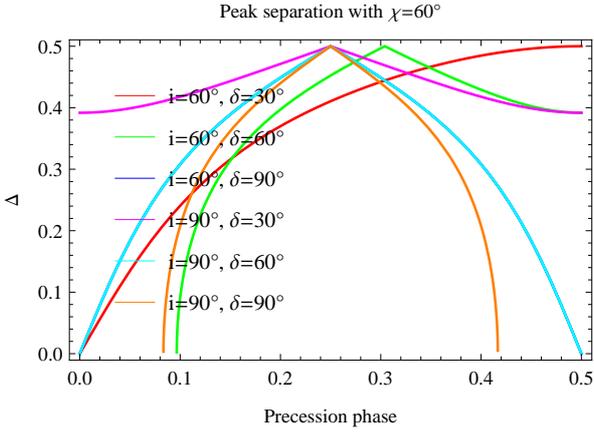} \\
 \caption{Theoretical predictions of the peak separation~$\Delta$ with respect to the precession phase~$\varphi$. The geometrical parameters are shown in the legend, they are $\delta=\{30^o, 60^o, 90^o\}$, $i=\{60^o, 90^o\}$ and $\chi=60^o$.}
 \label{fig:SeparationPicTheorie}
\end{figure}

It is possible to obtain a precise idea of the pulse-peak separation evolution with the geodetic precession phase according to a simple analytical expression. Indeed, it was shown in \cite{2011MNRAS.412.1870P} that the separation~$\Delta$ is given by
\begin{equation}
  \label{eq:SeparationPic}
  \cos(\pi\,\Delta) = |\cot \zeta \, \cot \chi|.
\end{equation}
From this derivation we conclude that the pulse profiles only expand or shrink if the viewing angle~$\zeta$ changes, $\chi$ being fixed by the neutron star magnetic field anchored in the crust. Inserting equation~(\ref{eq:zetaPrecession}) into equation~(\ref{eq:SeparationPic}), we immediately obtain the evolution of $\Delta$ with respect to the geodetic precession phase~$\varphi$. If no misalignment exists between orbital and spin momentum, that is for $\delta=0$, no precession is induced and the viewing angle remains constant such that $\zeta=i$. The same conclusion applies when the binary is seen face-on, that is with an orbital inclination~$i=0$. In both these cases, no geodetic precession is perceptible by a distant observer. No evolution of the light curves would be noticed. In that case, we have $\zeta=\delta$ and thus $\zeta$ is also constant in time. Therefore the only interesting cases are those with a misaligned spin axis, $\delta\neq0$, and a misaligned orbital momentum, $i\neq0$. However, for too small misalignment angles~$\delta$, the amplitude of the variation in the line of sight inclination would be too low to allow any detection. For concreteness, we chose parameters that deviate significantly from $0^o$ and take $\{\chi, i, \delta\} = \{30^o, 60^o, 90^o\}^3$. The viewing angle~$\zeta$ is shown in figure~\ref{fig:Zeta} for different obliquities~$\chi$, different inclination of the spin axis with respect to the orbital momentum~$\delta$, and different inclination of the orbital plane~$i$. The amplitude of the variation of $\zeta$ is largest when the coefficient in front of $\cos\varphi$ in equation~(\ref{eq:zetaPrecession}) is largest, that is, equal to unity. This occurs for $i=\delta=90^o$. It represents the most optimistic configuration for geodetic precession.
According to this evolution of $\zeta$ with respect to $\varphi$, we easily deduce the theoretical evolution of double pulse peak separation as described above. Examples are shown in figure~\ref{fig:SeparationPicTheorie}.
This is basically all the information we can extract from the analytical model. However, to achieve deeper knowledge of the influence of the geodetic precession on to the pulse profiles, we performed numerical simulations by integrating the synchrotron emissivity in the whole striped wind.

\subsection{Numerical simulations}

We computed a full set of light curves by changing the geometry of the binary system with respect to the observer. We now compare the previous expectations with full pulse-profile computations by direct numerical integration of the emissivity in the striped wind. The main results are described in this section.

\subsubsection{High-energy pulse profile evolution}

\begin{figure*}
 \centering
\begin{tabular}{cc}
\includegraphics[width=0.45\textwidth]{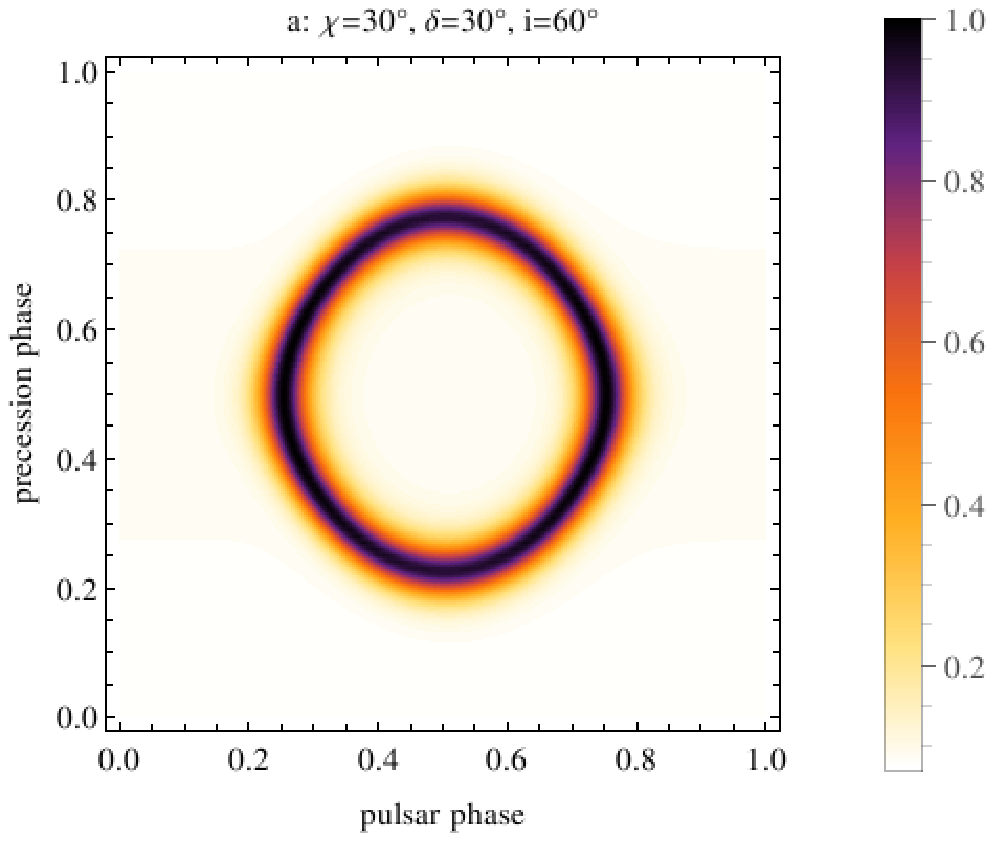} &
\includegraphics[width=0.45\textwidth]{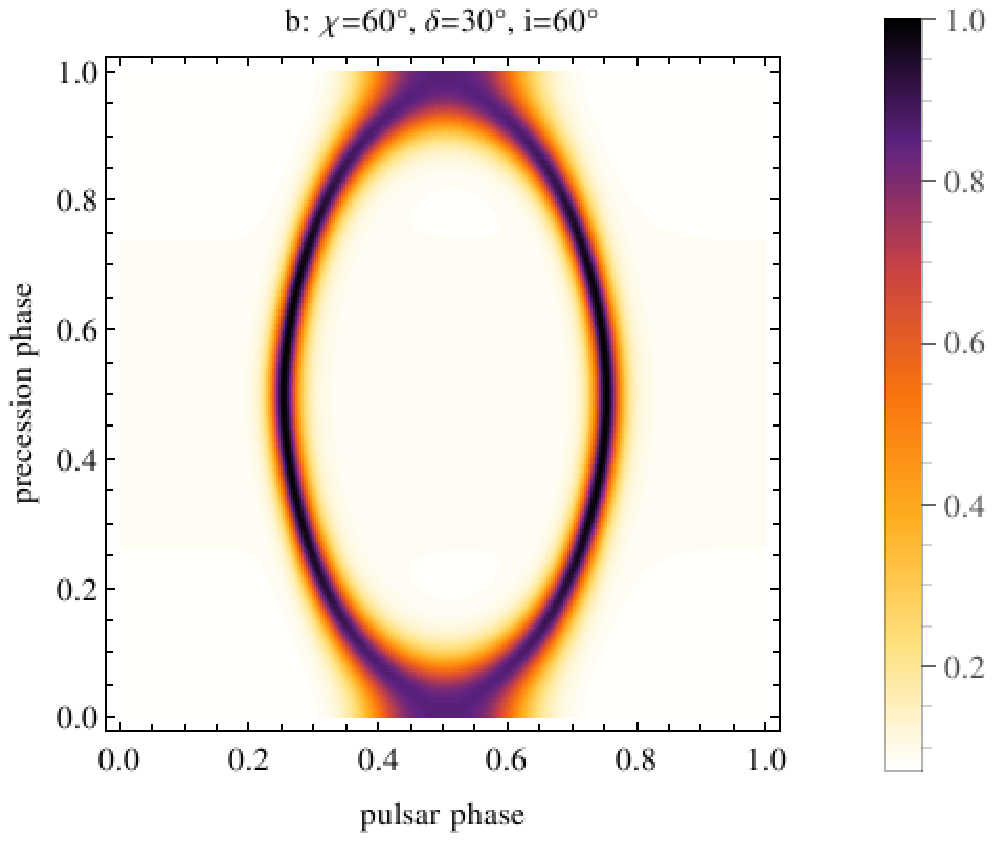} \\
\includegraphics[width=0.45\textwidth]{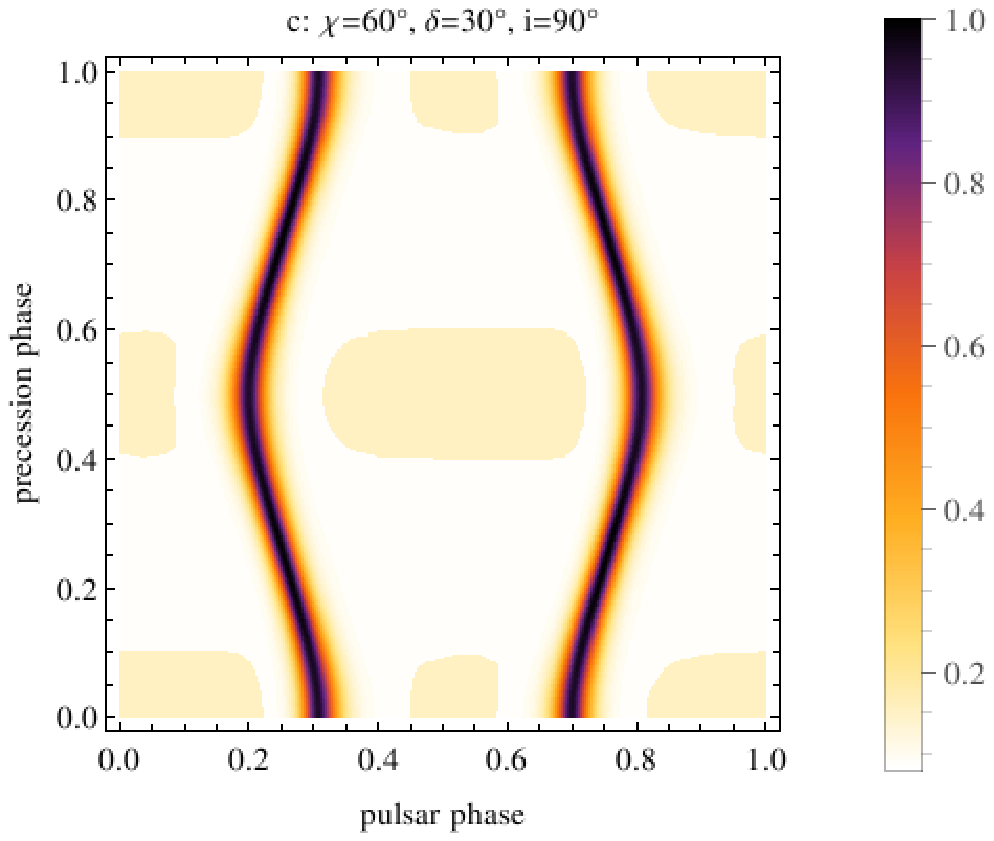} &
\includegraphics[width=0.45\textwidth]{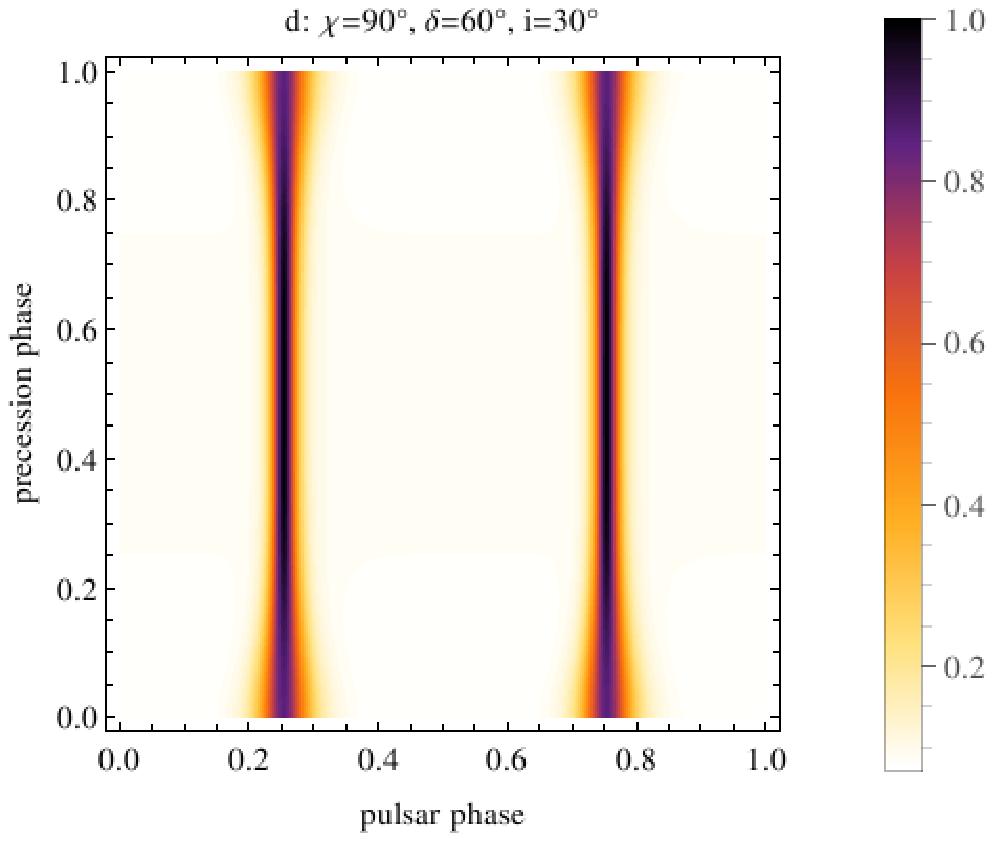}
\end{tabular}
 \caption{Evolution of the pulse profile with respect to the precession phase~$\varphi$ for several geometries of the binary pulsar. Panel~a corresponds to case~1, panel~b to case~2, panel~c to case~3, and panel~d to case~4.}
 \label{fig:CourbeLumiere}
\end{figure*}

Typical examples of high-energy pulse profile evolution are shown in figure~\ref{fig:CourbeLumiere}. As expected, the peak separation varies with the precession phase, the effect being dominant whenever we approach the configuration $i=\delta=90^o$. We described the different types of light curves in detail. First, for case~1 with $\chi=30^o$, the magnetic moment is weakly tilted with respect to the spin axis. We know that in such a configuration the line of sight of the observer can easily miss the striped region of the wind and therefore not detect any pulsation. This is indeed seen in figure~\ref{fig:CourbeLumiere}a. For $\delta=30^o$ and $i=60^o$, the amplitude of variation of $\zeta$ according to equation~(\ref{eq:zetaPrecession}) is large enough to observe different pulse profiles, from faint single pulses through bright single pulses up to well-separated double pulses. The line of sight inclination is generally bounded by $\delta\pm i$ and $\pi-(\delta\pm i)$. In the particular case~1, $30^o\leq\zeta\leq90^o$ thus $\zeta$ can be much lower than $(\pi-\chi)$~rad, which implies no pulsed emission, whereas for $\zeta=90^o$ we expect to see two symmetric pulses separated by~0.5 in phase. This is indeed shown in figure~\ref{fig:CourbeLumiere} with a transition between two extreme pulse profiles through a regime of faint or bright single pulses. Taking the same orbital parameters, but an obliquity $\chi=60^o$, the probability of missing the striped region becomes more unlikely because $\zeta$ can never be lower than $(\pi-\chi)$~rad so that we always see one or two pulses, as confirmed by inspecting panel~\ref{fig:CourbeLumiere}b. The light curve always shows strong pulsation switching from single to double pulse, but without disappearing. The line of sight remains always within the opening angle of the cone of the striped wind. In both cases, the separation between the peak is wide enough to be detected by any X-ray or gamma-ray instrument. In the third case (panel~\ref{fig:CourbeLumiere}c) the orbit is seen face-on with $i=90^o$, the other angles being the same as in case~2. The excursion in $\zeta$ is reduced to the range~$[60^o,120^o]$ and the observer line of sight always crosses the striped structure significantly. One therefore sees a double pulse shape, regardless of the precession phase of the pulsar. The light curves almost overlap and only a long-term data accumulation can distinguish different shapes. The variation in pulse intensity and separation decrease compared with previous cases, and a clear identification of geodetic precession from these light curves becomes difficult. The situation is even worse in the last configuration of an orthogonal rotator with $\chi=90^o$, $\delta=60^o$ and $i=30^o$, case~4 (panel~\ref{fig:CourbeLumiere}d). The peak separation remains always equal to 0.5 in phase in this special geometry. The only influence of geodetic precession is observed for the peak intensity and the pulse width. But as seen in the plot, the changes are not that obvious to detect, especially if data were contaminated by a large amount of noise. The general conclusion to be drawn from this study is that clearly recognizing a geodetic precession phenomenon depends on the orbital parameter of the system and the geometry of the pulsar. 
All the information about the peak separation, the peak highest intensity, and the pulse width are contained in these curves. We now extract these essential features to understand the effect of geodetic precession on the striped-wind pulsed emission.

\subsubsection{Peak separation of double pulses}

We discussed the peak separation~$\Delta$ from an analytical point of view. But do the numerical calculations agree with these predictions? To a very good accuracy, they do. The maxima of the two pulses can be tracked with the precession phase~$\varphi$. But first, we explain how we investigated the shape of the light curves obtained from our numerical simulations. To estimate the pulse width and related characteristics such as the highest intensity, we fitted the pulse profiles by either two Gaussian pulses given by
\begin{equation}
\label{eq:Gaussien}
 I_{\rm gau}(t) = I_0 + I_1 \, e^{-\left(\frac{\varphi-\varphi_1}{w_1}\right)^2} + I_2 \, e^{-\left(\frac{\varphi-\varphi_2}{w_2}\right)^2} ,
\end{equation}
or by two Lorentzian pulses such that
\begin{equation}
\label{eq:Lorentzien}
 I_{\rm lor}(t) = I_0 + \frac{I_1}{1+\left(\frac{\varphi-\varphi_1}{w_1}\right)^2} + \frac{I_2}{1+ \left(\frac{\varphi-\varphi_2}{w_2}\right)^2}.
\end{equation}
The fitting parameters are as follows:
\begin{itemize}
 \item $I_0$ is the background direct current (DC) component of the light curve.
 \item $I_1,I_2$ are the peak intensities within pulse~1 and~2 (DC component~$I_0$ subtracted).
 \item $\varphi_1,\varphi_2$ are the phases of the centre of each peak.
 \item $w_1,w_2$ are the characteristic widths of each pulse. We take these as a definition for the pulse width.
\end{itemize}
We therefore have seven parameters to fit for each model. $I_0$ is easily found to be close to the lowest intensity in the pulse profile, $(\varphi_1,\varphi_2)$ are around the phases of highest intensity given by $(I_1,I_2)$ respectively. These estimates are good first guesses for the fitting. 

\begin{figure*}
 \centering
\begin{tabular}{cc}
  \includegraphics[width=0.45\textwidth]{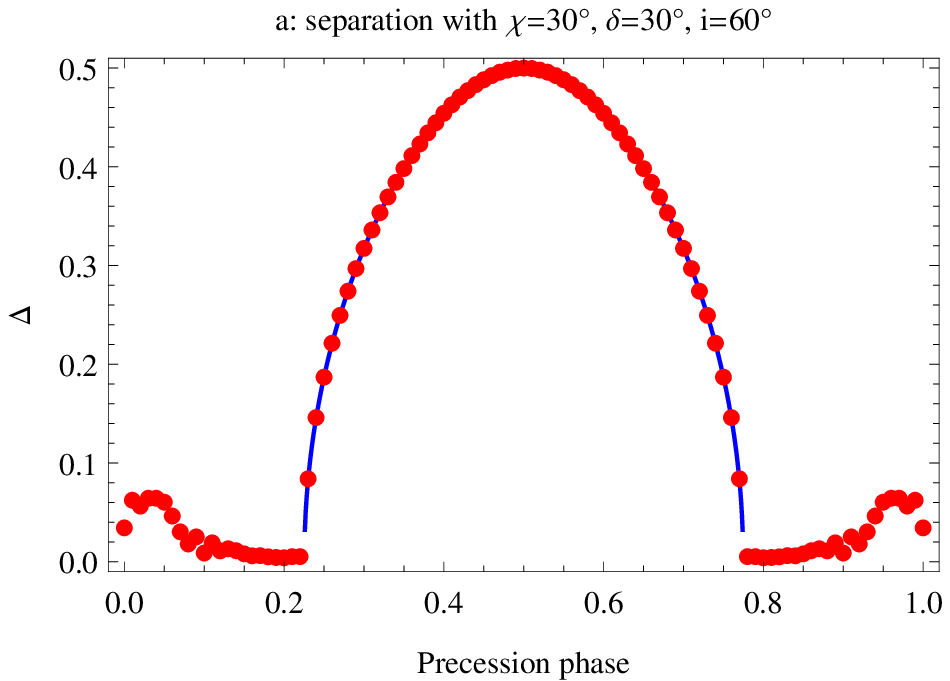} &
  \includegraphics[width=0.45\textwidth]{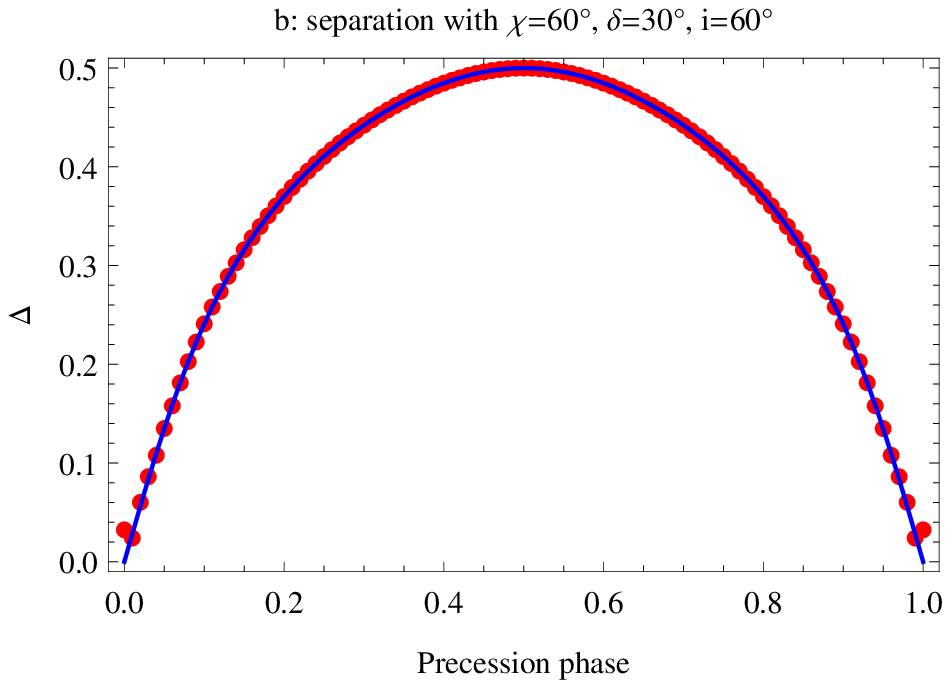} \\
  \includegraphics[width=0.45\textwidth]{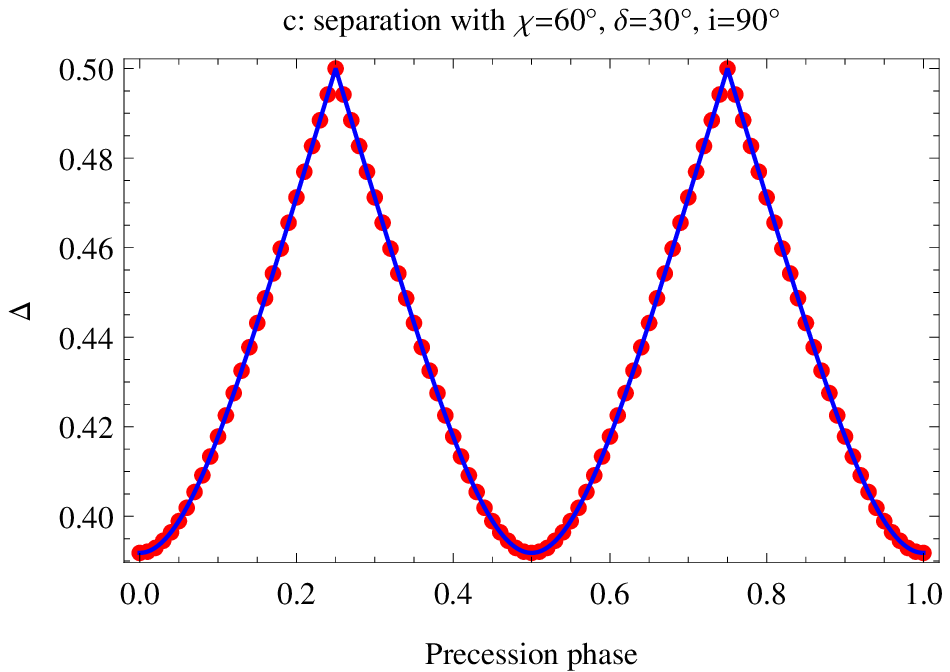} &
  \includegraphics[width=0.45\textwidth]{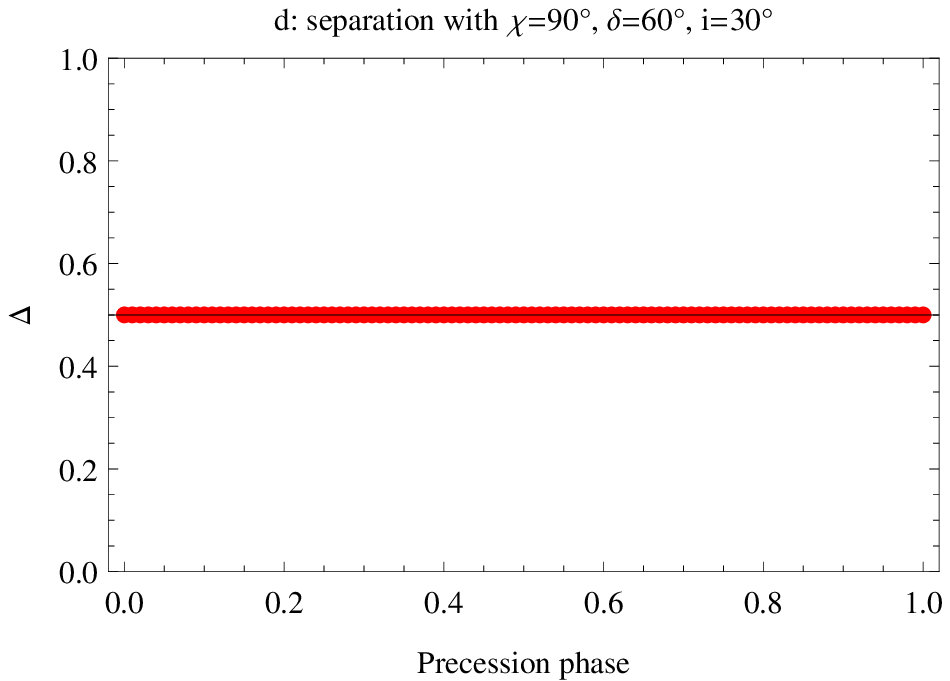} 
\end{tabular}
 \caption{Simulated evolution of the peak separation~$\Delta$ with respect to the precession phase~$\varphi$ compared with the predictions. The geometrical parameters are shown in the legend, they are $\delta=\{30^o, 60^o, 90^o\}$, $i=\{60^o, 90^o\}$ and $\chi=\{30^o, 60^o, 90^o\}$. They correspond to the same configuration as in figure~\ref{fig:CourbeLumiere}.}
 \label{fig:ComparaisonSeparation}
\end{figure*}

A few examples are shown in figure~\ref{fig:ComparaisonSeparation}. They correspond to the same parameters as those in figure~\ref{fig:CourbeLumiere}. These plots show that the analytical expression very well follows the more detailed computation of the pulse profiles presented in this section. Inspecting for instance  panel~\ref{fig:ComparaisonSeparation}a where both expectations overlap, its is impossible to distinguish the difference by eye as long as the two peaks are well separated. In phases where the two pulses strongly overlap, it is difficult to assess whether this is a large but single pulse or distinct two pulses. This explains the discrepancy between the two results in the wings in the left and right part of the plot. During these precession phases, the pulsed component is irrelevant, which means that extracting a peak separation becomes difficult. In the latter regime, the two peaks are not well separated, $\Delta<0.1$. The widest separation is obtained at a phase~$0.5$ and is equal to $0.5$. This is explained by the fact that for $\varphi=\pi$ the inclination of the line of sight is $\cos\zeta=\cos(\delta+i)$, thus $\zeta=\delta+i$ or $\zeta=\pi-(\delta+i)$. With the parameters of the first case we derive $\zeta=90^o$ at that phase. Therefore the separation should be exactly half of a period. The same conclusion applies to cases~2 and~4 because they possess the same sum $\delta+i=90^o$. For case~3, we derive $\delta+i=120^o$ therefore according to equation~(\ref{eq:SeparationPic}) we have $\Delta\approx0.392$ in agreement with panel~\ref{fig:ComparaisonSeparation}c. A separation $\Delta=0$ should be interpreted as the presence of at most one single pulse for instance like in case~1. Case~2 is a special configuration where one single pulse is only seen at phase $\varphi=0$. Case~4 is particularly simple because the separation remains equal to half a period whatever the inclination angle~$\zeta$. For the perpendicular rotator, we always expect $\Delta=0.5$ \citep{2011MNRAS.412.1870P}. At phase $\varphi=0$ the inclination of the line of sight is $\cos\zeta=\cos(\delta-i)$ thus $\zeta=\delta-i$ or $\zeta=\pi-(\delta-i)$. With the parameters of the first case we obtain $\zeta=30^o$ or $\zeta=150^o$, which implies no pulsed emission any more. For case~2, we also obtain $\zeta=30^o$ or $\zeta=150^o$ so that the observer looks at the edge of the striped region. He will not detect two pulses but only one. For case~3, $\zeta=60^o$ or $\zeta=120^o$, two pulses are still seen and $\Delta\approx0.392$ again. At the phase $\varphi=\pi/2$ and for $i=90^o$ we have $\zeta=90^o$ which explains the separation of $\Delta=0.5$ in panel~\ref{fig:ComparaisonSeparation}c. The same remarks holds for $\varphi=3\,\pi/2$.
In conclusion, the analytical formula given in the previous section is excellent to understand the peak separation evolution in the striped-wind scenario. We hope that these plots, if obtained from long-term observations, will help constraining the crucial angles involved in the pulsar high-energy emission geometry.

\subsubsection{Peak intensity evolution}

\begin{figure*}
 \centering
\begin{tabular}{cc}
  \includegraphics[width=0.45\textwidth]{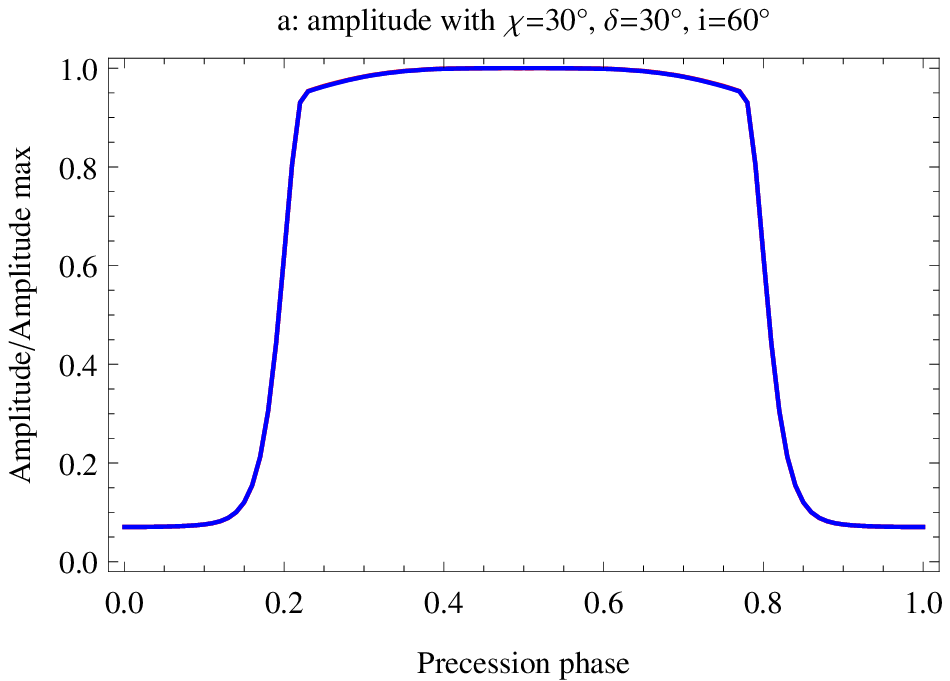} &
  \includegraphics[width=0.45\textwidth]{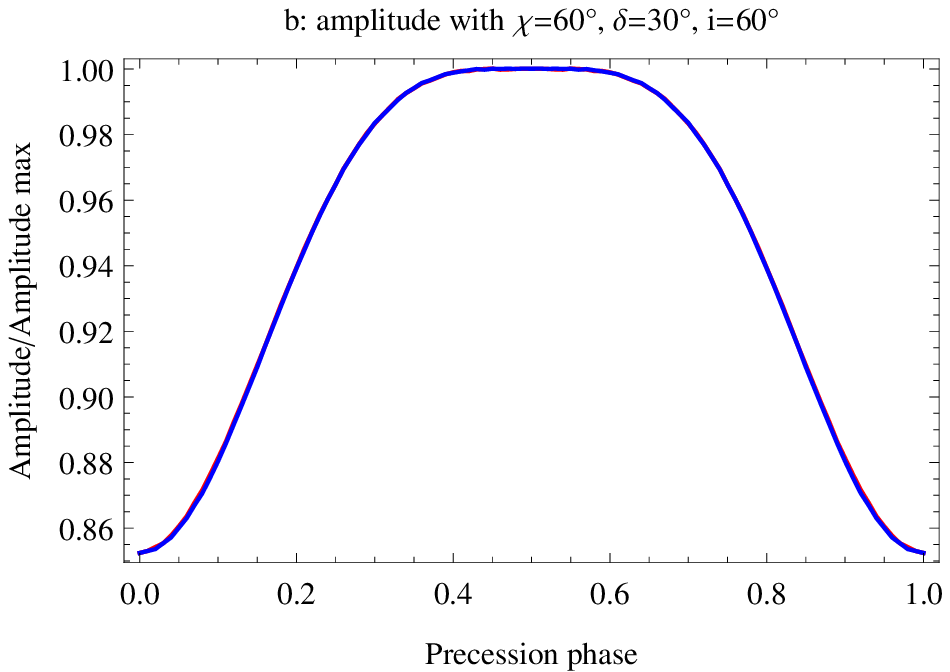} \\
  \includegraphics[width=0.45\textwidth]{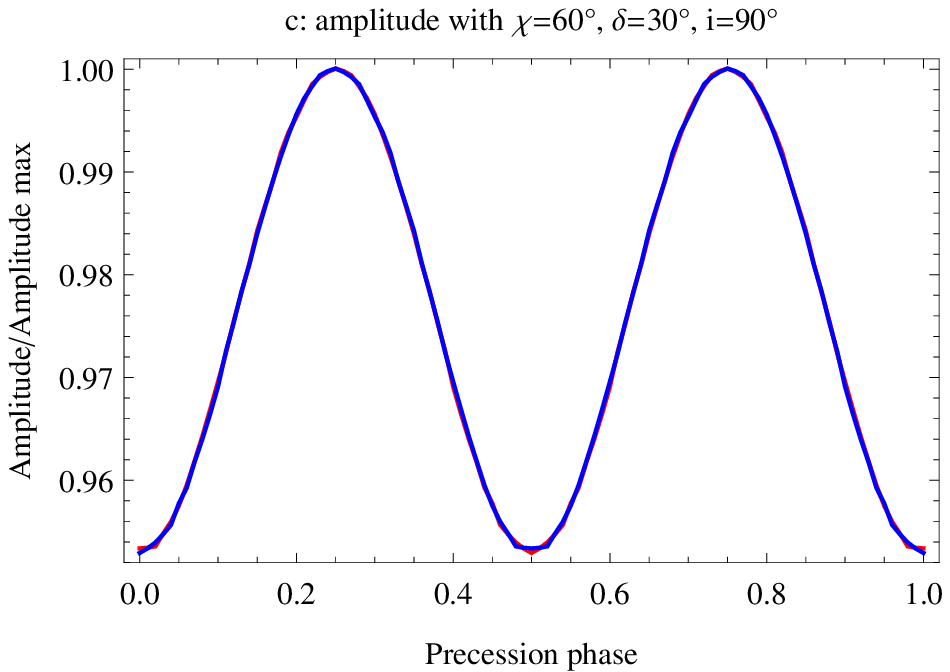} &
  \includegraphics[width=0.45\textwidth]{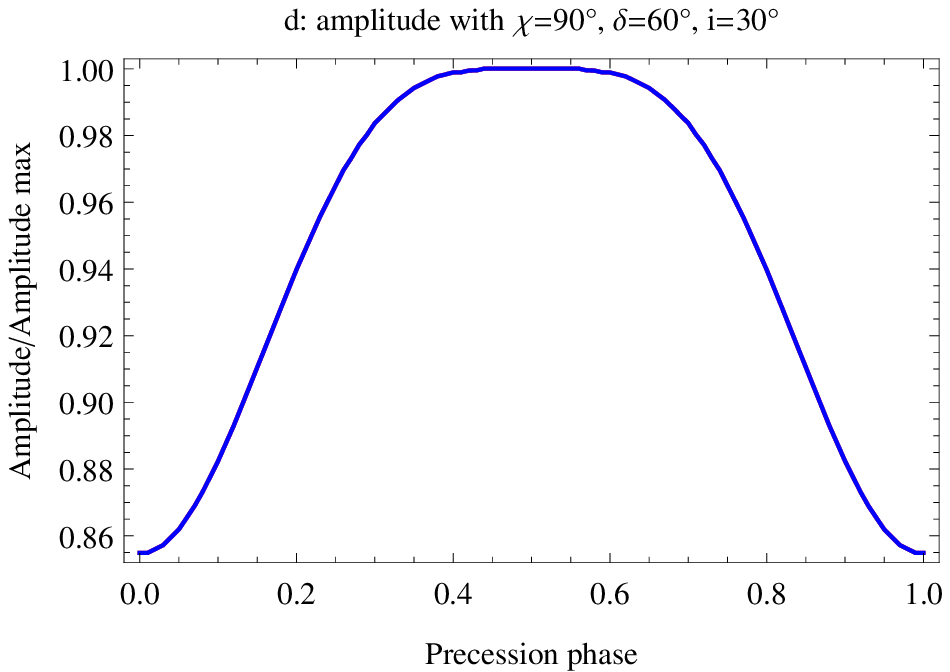} 
\end{tabular}
 \caption{Simulated evolution of the highest peak amplitude~$I_{\rm max}$ with respect to a full precession period with phase~$\varphi\in[0,2\,\pi]$. The geometrical parameters are shown in the titles. The two fitting models are shown, Gaussian in red and Lorentzian in blue. Both fits overlap almost perfectly. The highest intensity does not depend on the fitting profile, as expected. The intensity contrast can reach 10 for panel~a and be rather weak for other configurations, such as panel~c, which shows a variation of only 5\%.}
 \label{fig:Amplitude}
\end{figure*}

Tracking the peak intensity evolution on a theoretical basis as above is less obvious. Nevertheless, our model predicts a variation in the highest intensity of the two peaks during a period of precession. Depending on the geometrical parameters, the change can be drastic. In panel~\ref{fig:Amplitude}a, we have seen that the pulse disappears, which translates into a sharp decrease of the peak intensity before phase~0.2 and after phase~0.8. It reaches the lowest level, the DC component. Cases~2 and 4 show a less pronounced variation, only up to 15~percent, panels~\ref{fig:Amplitude}b,d. In case~3, the variation is only about 5~percent, but it reproduces a symmetric behaviour after phase~0.5 (panel~\ref{fig:Amplitude}c).
This demonstrates the variety of pulse shapes that are allowed by our simple geometric model. We finish our report with the evolution of the pulse width.

\subsubsection{Pulse width evolution}

\begin{figure*}
 \centering
\begin{tabular}{cc}
 \includegraphics[width=0.45\textwidth]{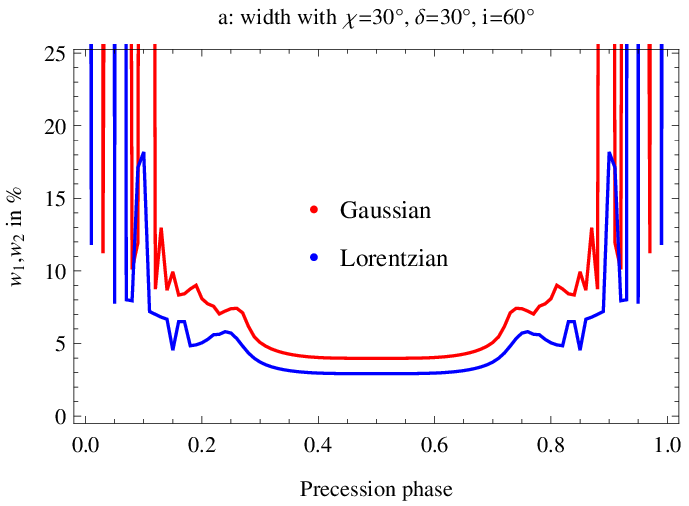} &
 \includegraphics[width=0.45\textwidth]{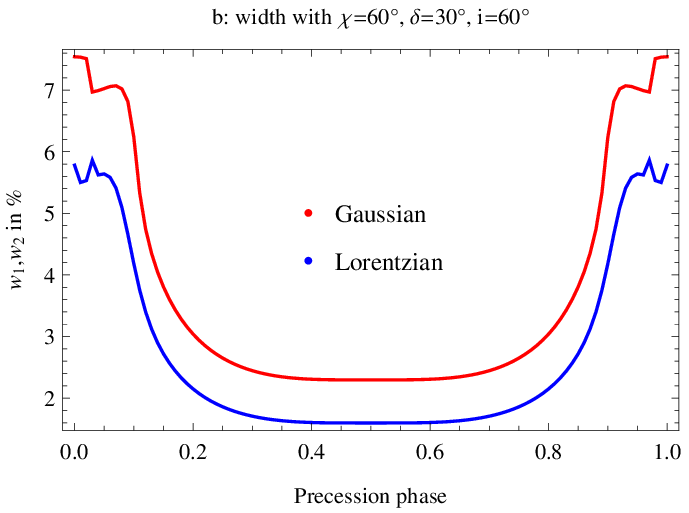} \\
 \includegraphics[width=0.45\textwidth]{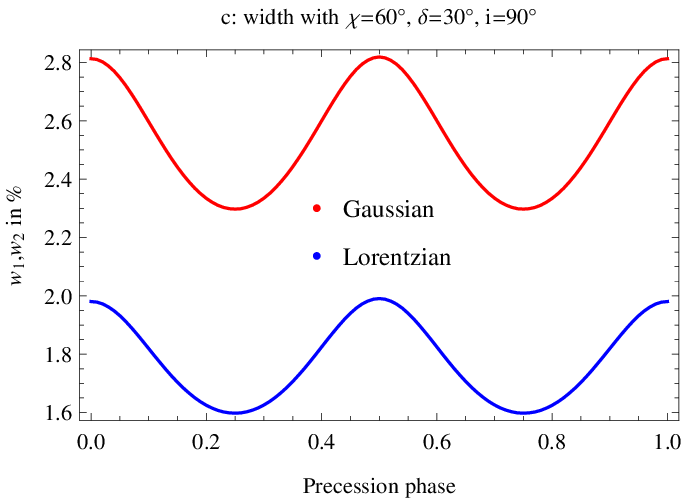} &
 \includegraphics[width=0.45\textwidth]{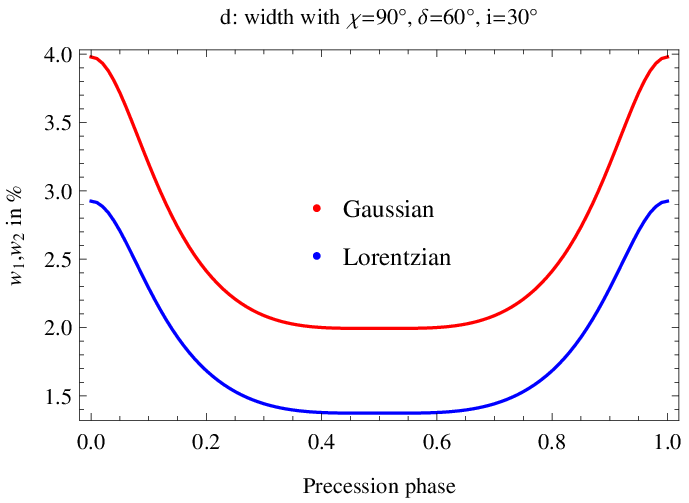}
\end{tabular}
 \caption{Evolution of the pulse width $(w_1,w_2)$ with respect to the precession phase~$\varphi$. The two fitting models are shown, Gaussian in red and Lorentzian in blue. For the first plot (panel~a) significant emission occurs only in the phase interval $\phi\in[0.2,0.8]$. This is because the line of sight does not intersect the high-energy beam anymore. The plot outside this range is therefore irrelevant.}
 \label{fig:Largeur}
\end{figure*}

According to the fit described in equations~(\ref{eq:Gaussien}) and (\ref{eq:Lorentzien}), the pulse width evolution can be tracked with respect to the geodetic precession angle. The results are reported in figure~\ref{fig:Largeur}. As a general remark, the Lorentzian fits always underestimate the width compared with the Gaussian fits. However, the evolution with precession phase remains the same in both models. When the separation is widest and equal to $\Delta=0.5$, we observe a minimum in the pulse width, regardless of the angles. A decrease in separation of the peaks is always accompanied by an increase in the pulse width. Note that for case~1 (panel~\ref{fig:Largeur}a) the values of the width are meaningless before $\varphi=0.2$ and after $\varphi=0.8$ because there is no significantly pulsed emission at those phases. The strongest fractional changes are seen in case~2 (panel~\ref{fig:Largeur}b) the weakest changes correspond to case~3 (panel~\ref{fig:Largeur}c).

\section{DISCUSSION}
\label{sec:Discussion}

So far, we discussed the influence of geodetic precession on high-energy light curves in the pulsar striped-wind scenario in general. But will it be possible in the future to detect these effects in X-rays or gamma-rays? Following \cite{2010arXiv1008.5032K}, any observation of spin precession in a binary pulsar requires
\begin{enumerate}
 \item a misalignment~$\delta$ between the pulsar spin axis and the orbital angular momentum.
 \item a precession period~$2\,\pi/\Omega_{\rm p}$ sufficiently short to detect a variation within a reasonable observation window, a few years or a few decades at most.
 \item a long-term regular observation of the pulsed emission accumulating many years of data.
\end{enumerate}
He also listed possible target double neutron star systems classified according to decreasing precession rate~$\Omega_{\rm p}$. We review some of these interesting candidates.

The list of target pulsars is summarized in Table~\ref{tab:Echantillon}. They all possess relatively high geodetic precession rates of about $1^o$/yr. Those with the highest spin precession rate, even if not detected in high-energy yet, might have a chance to suddenly appear as a consequence of this spin precession. This would be a valuable information about the high-energy emission processes occurring in pulsar winds and magnetospheres. The figure-of-merit, $\sqrt{\dot E}/d^2$, as defined by \cite{2013ApJS..208...17A}, is also shown. We expect pulsars with $\sqrt{\dot E}/d^2\gtrsim0.3\times10^{13}~\sqrt{\textrm{ W}}/kpc^2$ to be detectable by Fermi/LAT if the line of sight of the observer crosses the striped part of the wind. PSR~J0737-3039A has indeed been seen by Fermi/LAT, but the geometry of the spin precession is not favourable. In our opinion, the best candidates for which we might expect to see an evolution of the high-energy pulse profile are PSR~J1906+0746 and PSR~J1141-6545. They possess a high geodetic spin precession with almost orthogonal axis and a reasonable figure-of-merit.

\begin{table*}
\centering
\begin{tabular}{llllllllll}
\hline
 Pulsar & P & $\dot E$ & $d$ & $\Omega_{\rm p}$ & $\chi$ & $i$ & $\delta$ & $\sqrt{\dot E}/d^2$ & Reference \\ 
  & (ms) & $(10^{26}\textrm{ W})$ & (kpc) & (deg/yr) &  &  & & $(10^{13}\,\sqrt{\textrm{ W}}/kpc^2)$ &  \\ 
\hline
\hline
 J0737-3039A & 22.7 &    5.9 & 1.1 & 4.8 &  90 & 90 & $\lesssim6$ & 2.0 & \cite{2013ApJ...767...85F} \\ 
 J0737-3039B & 2773 & 0.0017 & 1.1 & 5.1 &  70 & 90 & 138 & 0.03 & \cite{2010ApJ...721.1193P} \\ 
 J1906+0746  &  144 &    270 & 4.5 & 2.2 &  81 & 42-51 & 89 & 0.81 & \cite{2013IAUS..291..199D} \\
 B1913+16    &   59 &    1.7 & 7.1 & 1.2 &  47 & 157 & 21 & 0.026 & \cite{2002ApJ...576..942W} \\
 B1534+12    & 37.9 &    1.8 & 1.0 & 0.5 & 110 &  78 & 30,100? & 1.3 & \cite{2003ApJ...589..495K} \\
 J1141-6545  &  394 &    2.8 & 3.0 & 1.4 & 160 &  73 & 93 & 0.19 & \cite{2010ApJ...710.1694M} \\
\hline
\end{tabular}
\caption{Some characteristics of the double neutron star systems with their approximate orbital parameters. Angles are given in degrees according to the fits given in the corresponding references. We also computed the figure-of-merit as defined by \cite{2013ApJS..208...17A}. Distance are from the ATNF catalogue.}
\label{tab:Echantillon}
\end{table*}

We discuss the results for each system listed in Table~\ref{tab:Echantillon} at some length. We present the evolution of the light curve for a full geodetic precession, the variation in the highest intensity of one peak, a sample of light curves at different precession phases, and the double-peak separation for a full geodetic precession.

\subsection{PSR~J0737-3039}

\begin{figure*}
 \centering
\begin{tabular}{cc}
 \includegraphics[width=0.45\textwidth]{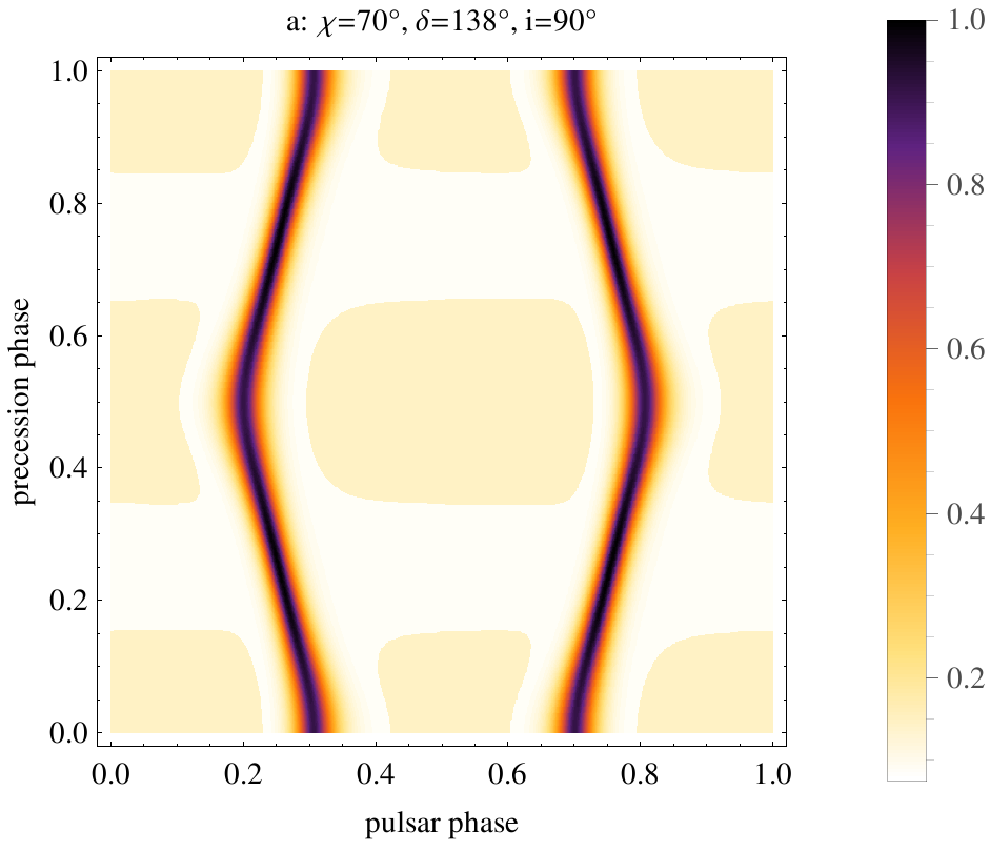} &
 \includegraphics[width=0.45\textwidth]{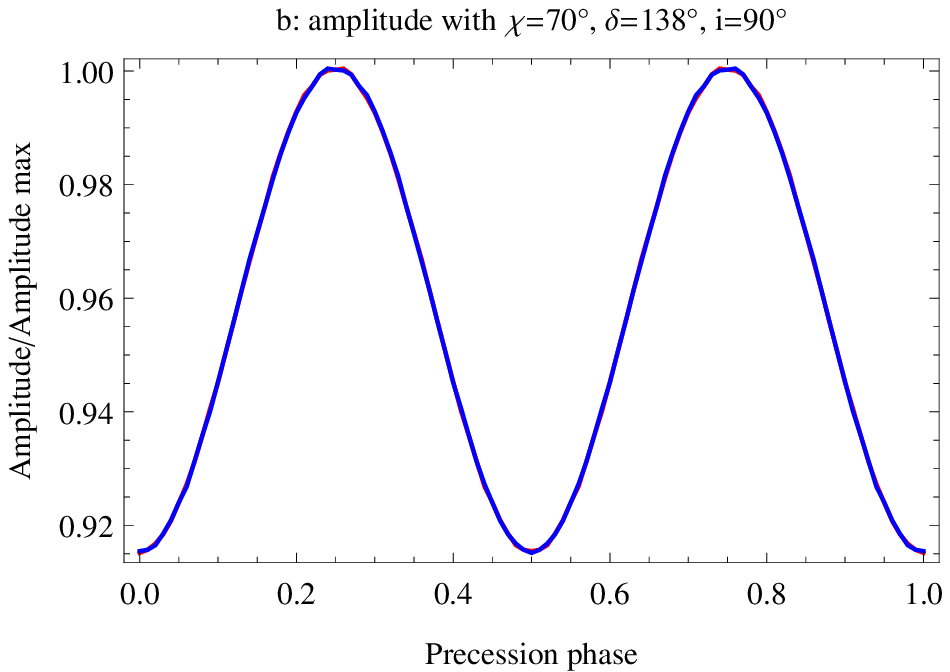} \\
 \includegraphics[width=0.45\textwidth]{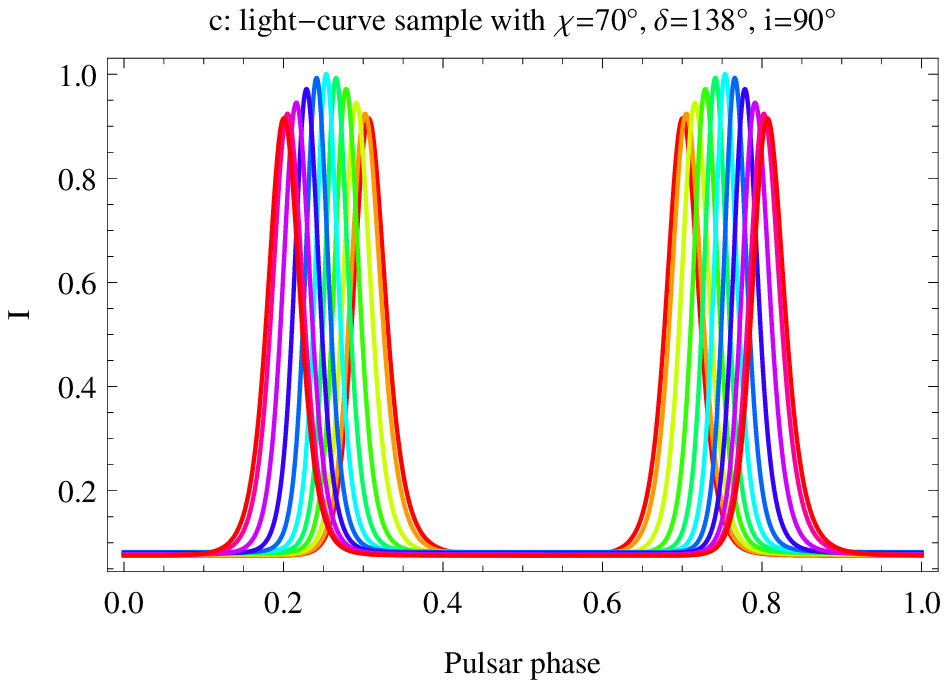} &
 \includegraphics[width=0.45\textwidth]{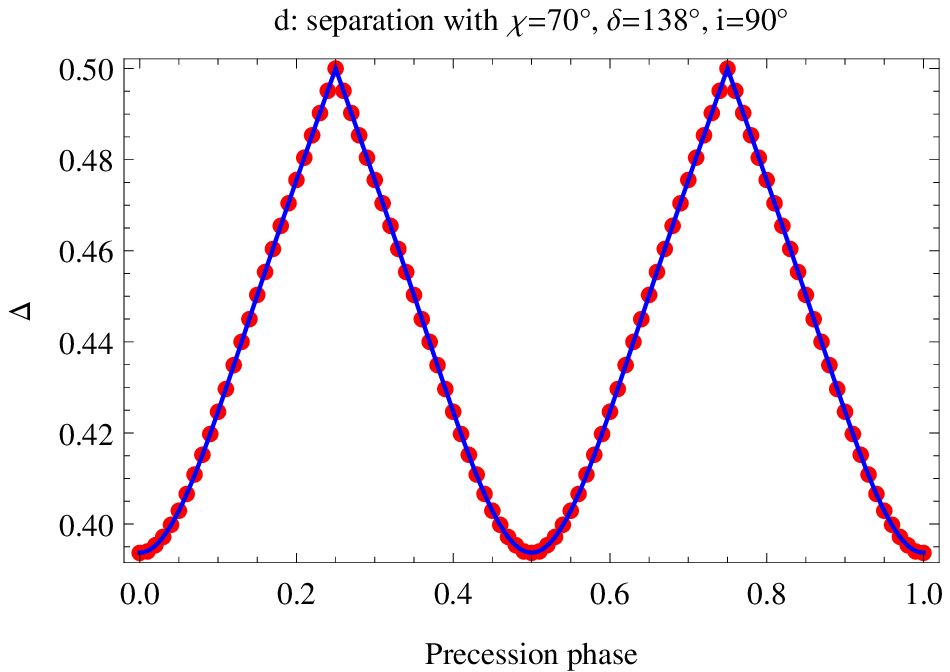}
\end{tabular}
 \caption{Detailed analysis of the evolution of the pulsed emission from PSR~J0737-3039B. From left to right and top to bottom we show the pulse profiles, the amplitude of the peak intensity, the width of the pulses, and the double-peak separation. Note the different x-label, pulsar phase for the light curves but precession phase for the other plots.}
 \label{fig:PSRJ0737}
\end{figure*}

Our present study was motivated by the discovery of pulsed gamma-ray emission from the double pulsar, PSR~J0737-3039 \citep{2013ApJ...768..169G}. The high precession rate of pulsar~A, which has a period of 22.7~ms, combined with pulsed emission seen by Fermi/LAT opens up the possibility of observing a secular evolution of the pulse profiles within a few years. Unfortunately, for this system, the misalignment between the pulsar~A spin axis and the orbital angular momentum induces only a weak effect that is not detectable in the pulse profile with our current technology. But it is conceivable that in the future such small variations might be seen by more sensitive telescopes which would eventually mean that we might be able to distinguish between competing high-energy pulsar models. \cite{2013ApJ...767...85F} reported a misalignment at most equal to $\delta\approx6^o$ assuming emission coming from both magnetic poles. Moreover, because the flux intercepted by Fermi/LAT is faint, there is little hope to see such fine characteristics in the light curves with current satellites, even after long integration times. We still await new technology with better sensitivity. Nevertheless, pulsar~B, with a period of 2.7~s, is an interesting object because according to \cite{2010ApJ...721.1193P} and \citet{2008Sci...321..104B}, its misalignment is about $\delta\approx20^o$. It is possible that this pulsar will show a pulsed high energy component in the next decade if this beam will point toward Earth. Nothing forbids the appearance of another component in the pulse profile caused by geodetic precession. Radio pulsation evolves and sometimes disappears, meanwhile, a high-energy component could appear because of the changing viewing geometry. To predict the shape of its light curve, we therefore computed its pulsed emission component. Its high-energy pulse profile is shown in panel~\ref{fig:PSRJ0737}a. The double-peak structure is persistent, but unfortunately, it has not been seen in high energy. Moreover, the change in peak intensity (panel~\ref{fig:PSRJ0737}b) and peak separation (panel~\ref{fig:PSRJ0737}d) is weak. Fluctuations in the light curves, as presented in panel~\ref{fig:PSRJ0737}c, remain in a narrow range. Its low figure-of-merit, Table~\ref{tab:Echantillon}, makes it invisible for Fermi/LAT. This means that we have to look for other promising candidates that might emit in X-rays and/or gamma-rays, and have appreciable precession rates. Fortunately, such double neutron star systems exist.

\subsection{PSR~J1906+0746}

\begin{figure*}
 \centering
\begin{tabular}{cc}
 \includegraphics[width=0.45\textwidth]{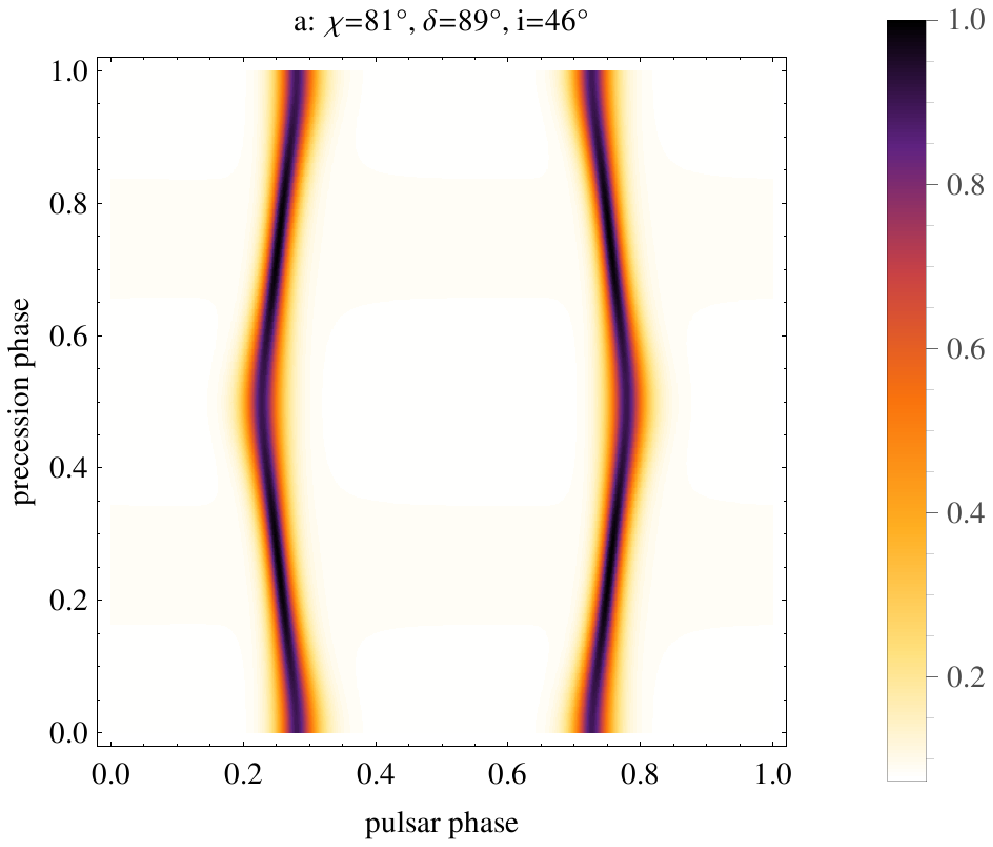} &
 \includegraphics[width=0.45\textwidth]{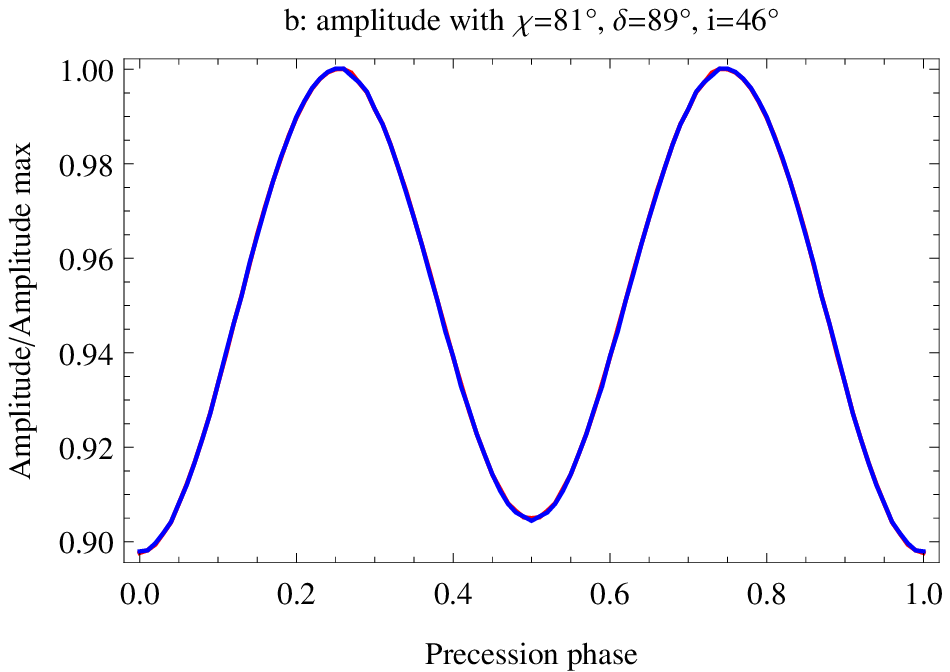} \\
 \includegraphics[width=0.45\textwidth]{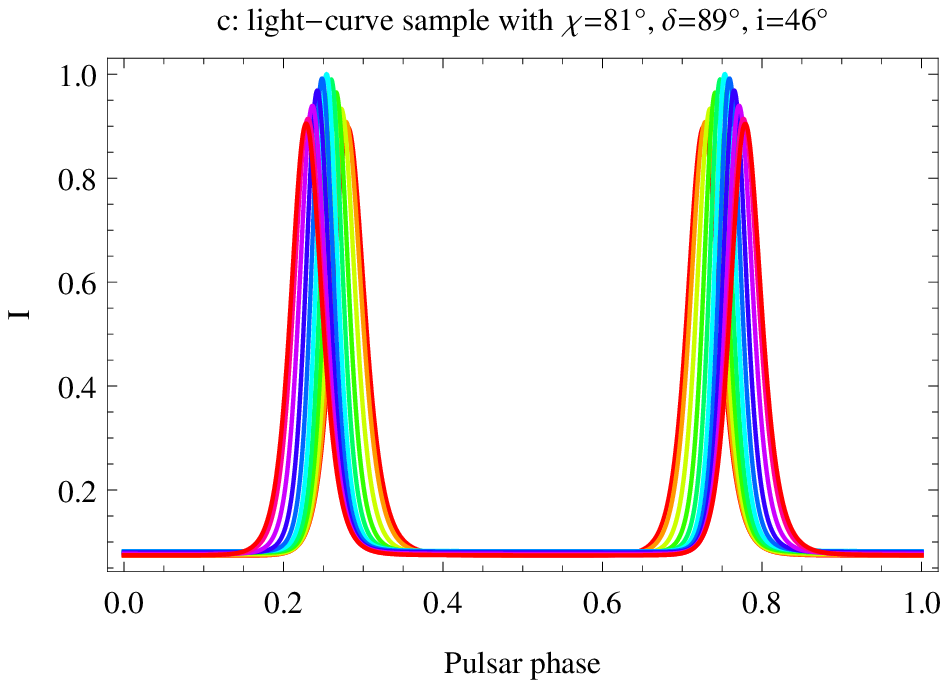} &
 \includegraphics[width=0.45\textwidth]{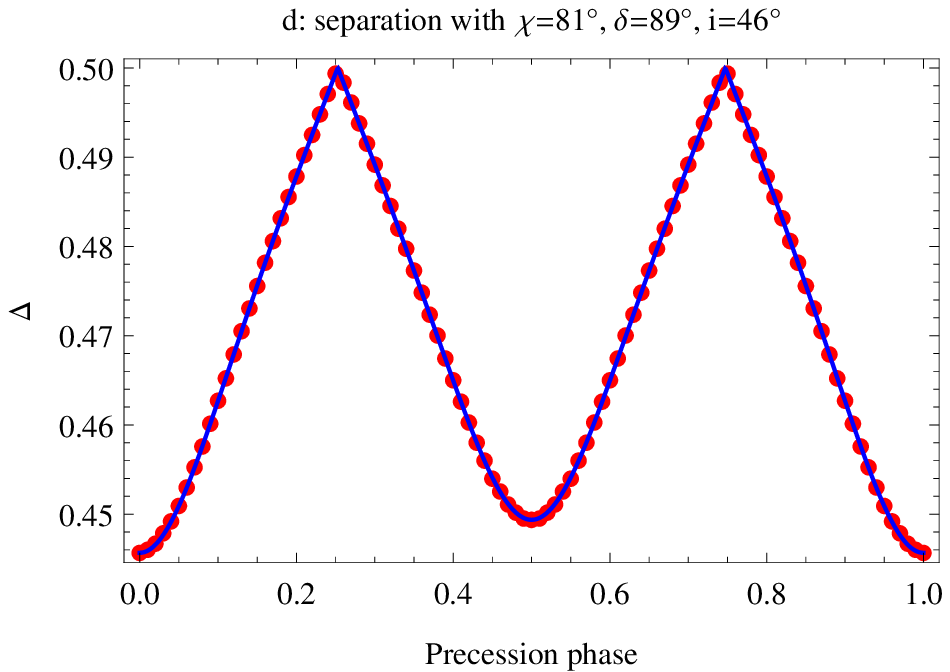}
\end{tabular}
\caption{Detailed analysis of the evolution of the pulsed emission from PSR~J1906+0746. From left to right and top to bottom we show the pulse profiles, the amplitude of the peak intensity, the width of the pulses, and the separation of the two peaks.}
 \label{fig:PSRJ1906}
\end{figure*}

PSR~J1906+0746 is a young pulsar with a period of 144~ms with an eccentricity $e=0.09$ and an orbital period of 4~hr. \cite{2013IAUS..291..199D} determined the geometry of this pulsar using the rotating vector model \citep{1969ApL.....3..225R}. They also produced maps of its radio beam. They found a nearly orthogonal rotator with $\chi\approx80^o$ and a misalignment of $\delta\approx89^o$, but with large error bars. PSR J1906+0746 is not reported in the second Fermi pulsar catalogue of \cite{2013ApJS..208...17A}. Because it is an almost orthogonal rotator, it is very unlikely that pulsed emission from the striped wind (which covers almost the $4\pi$~sr in that case) would remain undetectable if it were a gamma-ray pulsar. Therefore assumed that PSR J1906+0746 emits high-energy photons at a level insufficient to be detectable by space telescopes and inspect its light curve. A double-peak profile can be observed at any precession phase (panel~\ref{fig:PSRJ1906}a). As for the double pulsar, the changes in the high-energy features (panel~\ref{fig:PSRJ1906}b,c,d) remain within a restricted range that probably cannot be readed by current instruments. Therefore we must conclude that PSR~J1906+0746, which possesses a high spin-down luminosity of $3\times10^{28}$~W, if it were a gamma-ray pulsar, should be detected by the Fermi/LAT instrument, its figure-of-merit is appreciable. A well-separated double-pulse light curve should be observed at any spin precession phase. But its efficiency might be too low, or the fitting to the dipolar field by the rotating vector model is irrelevant. These could be some reasons for a non-detection.

\subsection{PSR~B1913+16}

\begin{figure*}
 \centering
\begin{tabular}{cc}
 \includegraphics[width=0.45\textwidth]{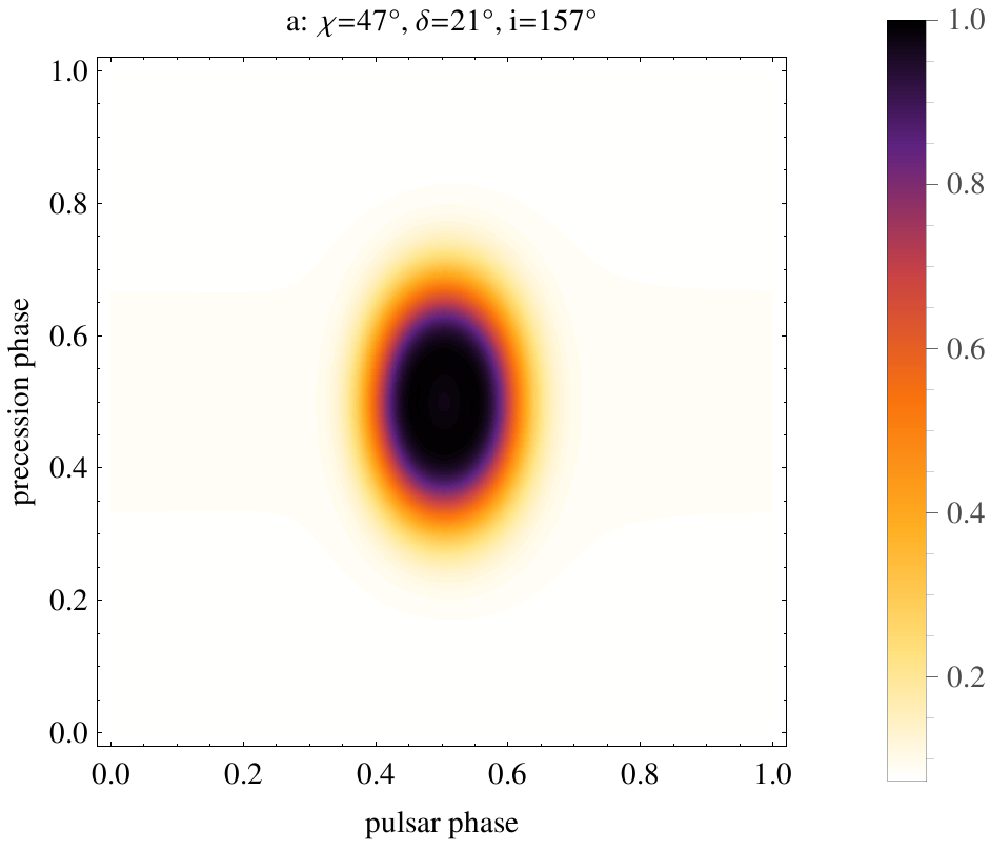} &
 \includegraphics[width=0.45\textwidth]{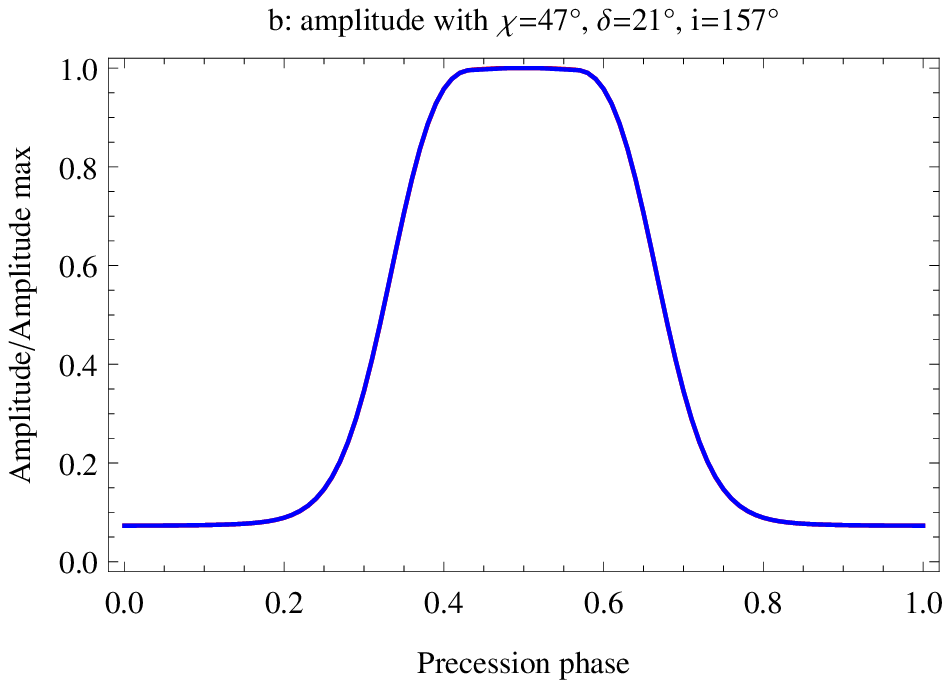} \\
 \includegraphics[width=0.45\textwidth]{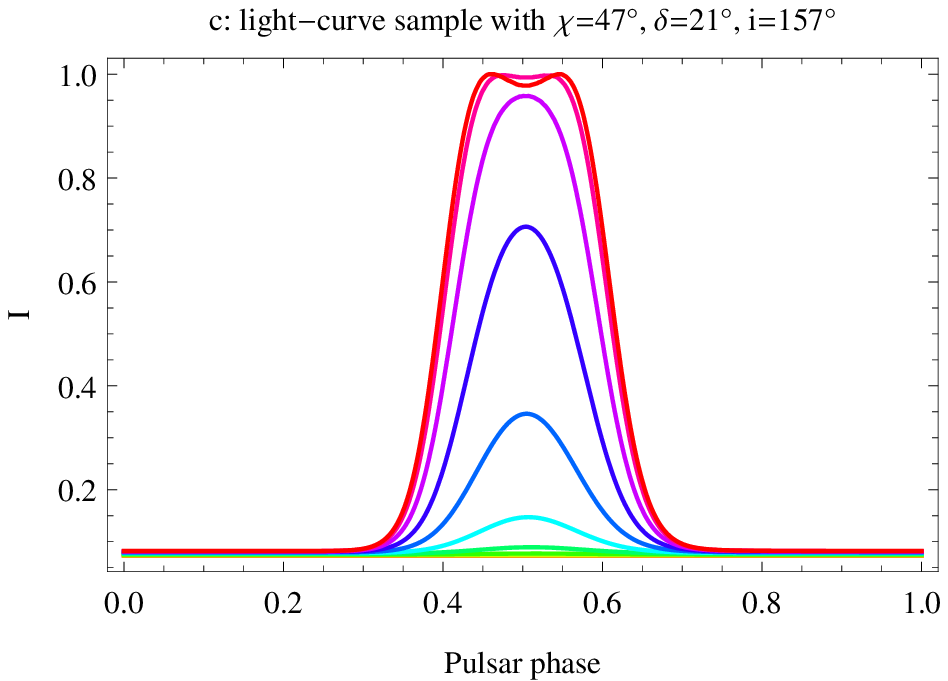} &
 \includegraphics[width=0.45\textwidth]{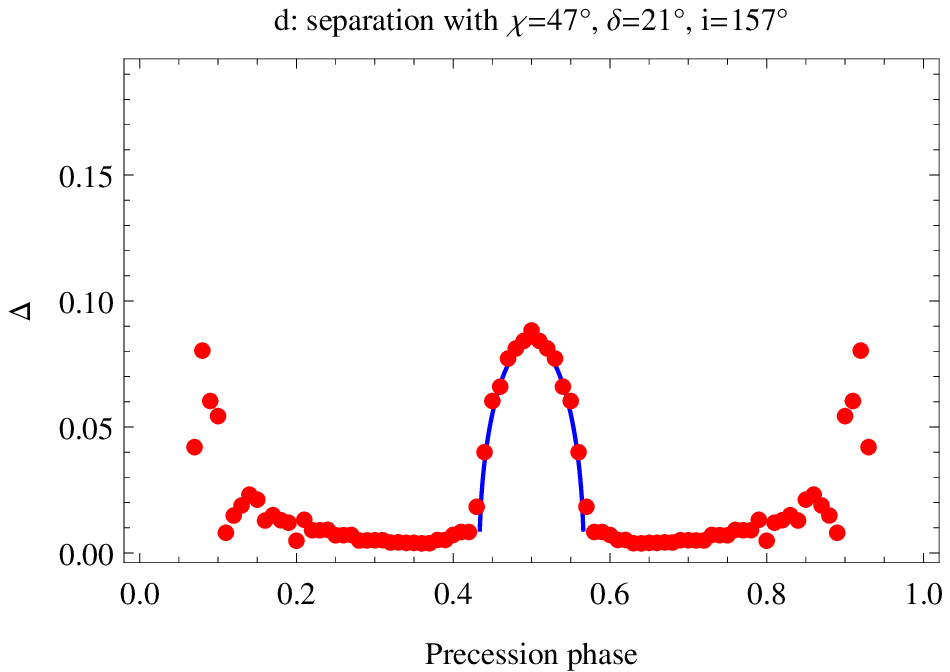}
\end{tabular}
 \caption{Detailed analysis of the evolution of the pulsed emission from PSR~B1913+16. From left to right and top to bottom, we show the pulse profiles, the amplitude of the peak intensity, the width of the pulses, and the separation of the two peaks.}
 \label{fig:PSRB1913}
\end{figure*}

PSR~J1913+16, also known as the Hulse-Taylor pulsar, with a period of 59~ms, was the first ever detected double neutron star system. Based on the radio pulse profile and polarization at 21~cm that were observed for 20~years, \cite{2002ApJ...576..942W} found the best fit of this pulsar with $\chi=158^o$, $i=47^o$, and $\delta=21^o$. No X-rays or gamma-rays have been reported from this neutron star so far. Moreover, the low inferred value of $\delta$ renders a detection of pulse profile variation difficult. However, our computations show that depending on the true spin precession phase, the change in the pulse profile can be dramatic, see panel~\ref{fig:PSRB1913}a. It is possible that this system is currently in a precession phase configuration that is unfavourable to any high-energy pulsation detection. It could for instance that within a few decades the highest intensity in the light curves increases dramatically, as shown in panel~\ref{fig:PSRB1913}b. The corresponding evolution of the pulse profile and the separated double peak are depicted in panel~\ref{fig:PSRB1913}c and panel~\ref{fig:PSRB1913}d. Another possible explanation for the non-detection is that the dipole model fails for this binary therefore the geometric parameters should be taken with caution as explained by \cite{1990ApJ...349..546C}. This might also explain that no high-energy radiation has been detected. Even if it is a gamma-ray pulsar, its low figure-of-merit as described in the Fermi/LAT gamma-ray pulsar catalogue leaves only little opportunity of measuring a gamma-ray flux in the near future.

\subsection{PSR~B1534+12}

\begin{figure*}
 \centering
\begin{tabular}{cc}
 \includegraphics[width=0.45\textwidth]{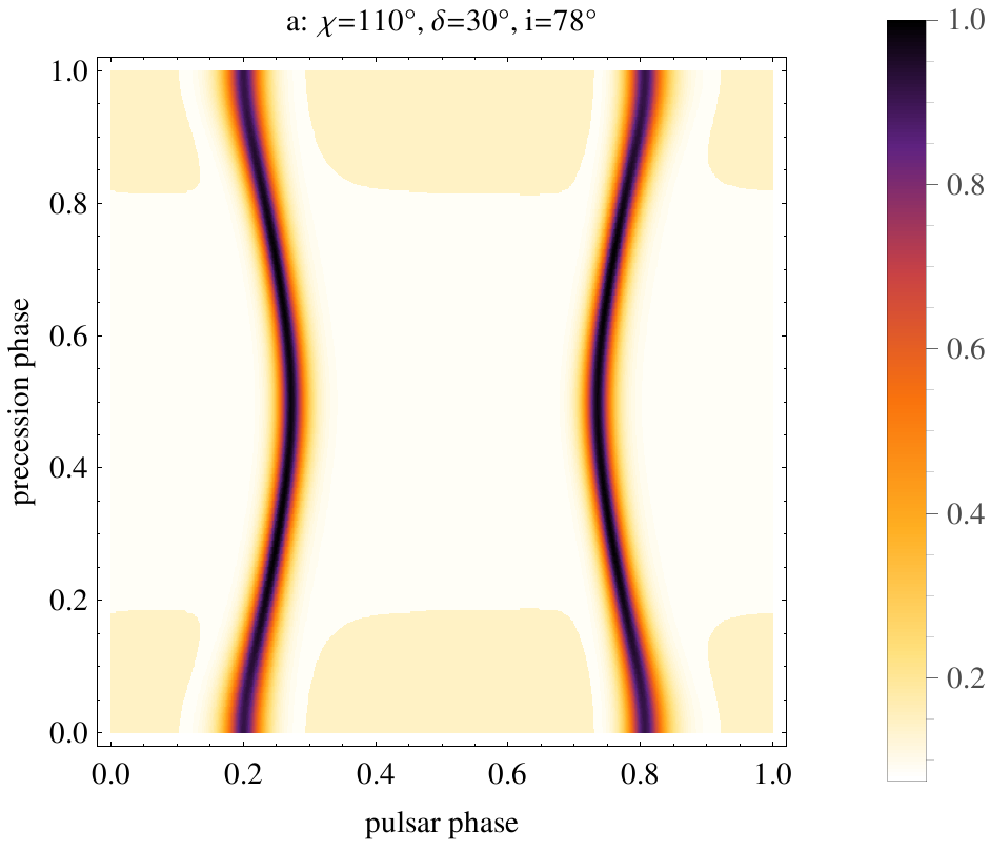} &
 \includegraphics[width=0.45\textwidth]{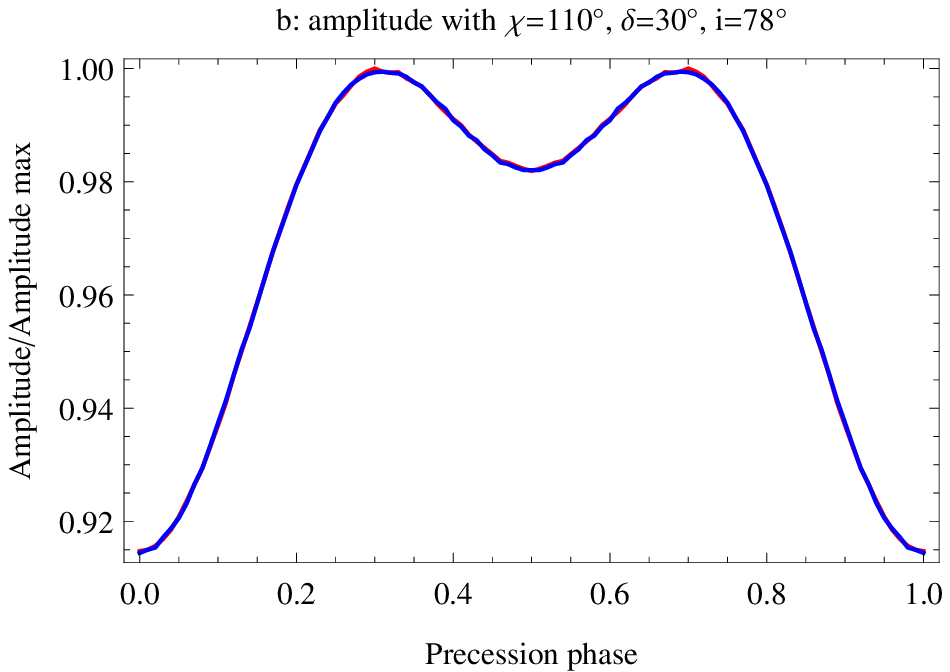} \\
 \includegraphics[width=0.45\textwidth]{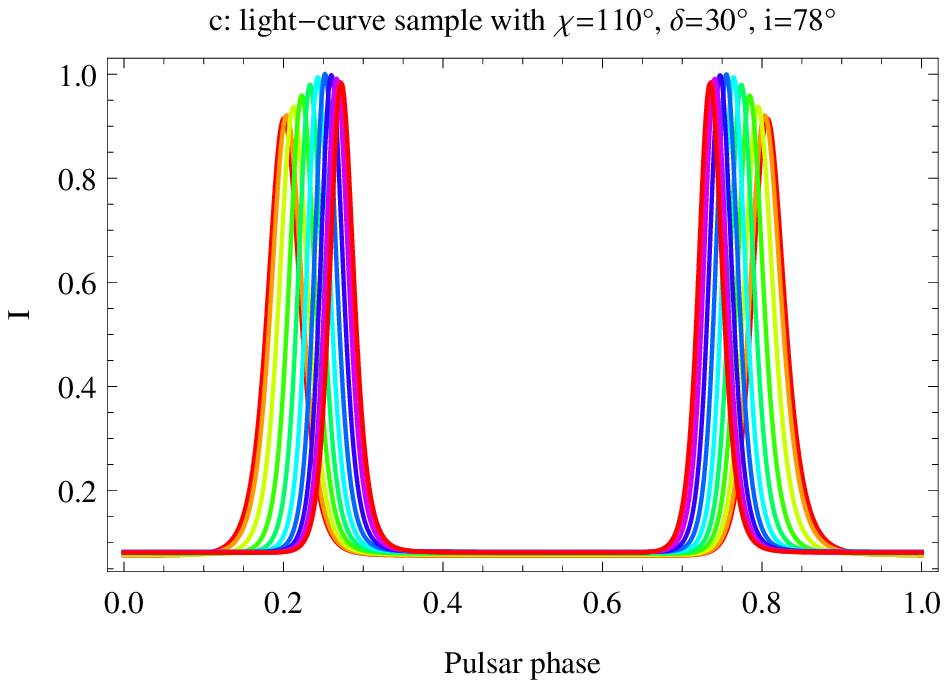} &
 \includegraphics[width=0.45\textwidth]{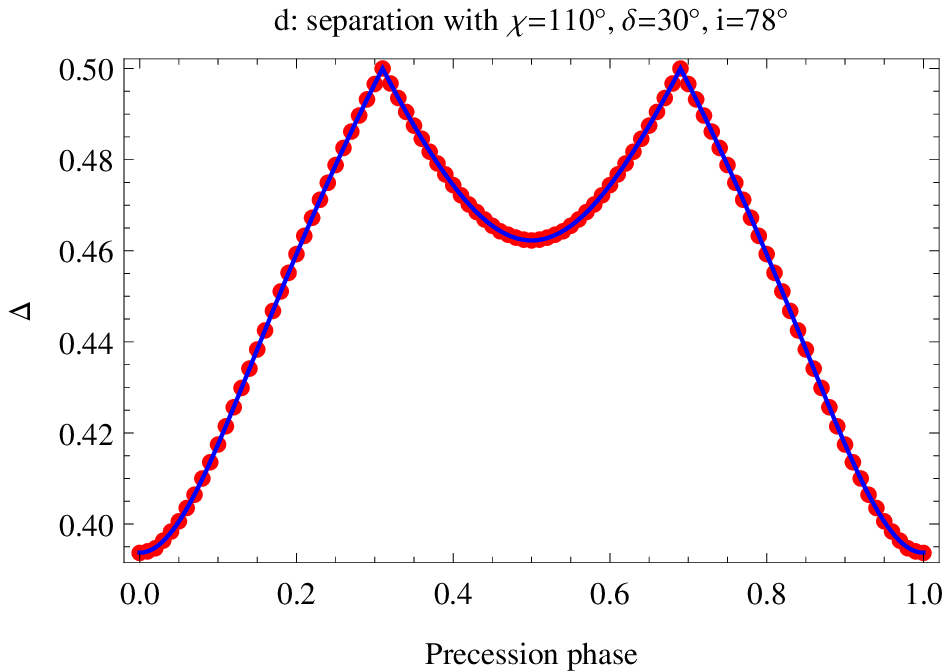}
\end{tabular}
 \caption{Detailed analysis of the evolution of the pulsed emission from PSR~B1534+12. From left to right and top to bottom, we show the pulse profiles, the amplitude of the peak intensity, the width of the pulses, and the separation of the two peaks.}
 \label{fig:PSRB1534}
\end{figure*}

PSR~J1534+12 is a millisecond pulsar with a period of 3.79~ms orbiting another neutron star \citep{1991Natur.350..688W}. \cite{2003ApJ...589..495K} were unable to constrain the geodetic precession rate because they lacked data on this pulsar. Nevertheless, \cite{2004PhRvL..93n1101S} later deduced two possible geometries for this system. Here we show results for one possibility: $\chi=110^o$, $\delta=30^o$, and $i=78^o$, see also \cite{2005ApJ...619.1036T}. Its precession rate of $0.5^o$/yr makes it difficult to expect significant changes in the light curves in the near future. We are nevertheless able to predict possible future detections in X-rays and/or gamma-rays. Expectations about its light curve are given in figure~\ref{fig:PSRB1534} for $\delta=30^o$. Moreover, so far there is no gamma-ray detection reported in the second Fermi/LAT gamma-ray pulsar catalogue. However, PSR~J1534+12 is an X-ray emitter \citep{2006ApJ...646.1139K} with orbital variation \citep{2011ApJ...741...65D}. The spin-down luminosity of about $2\times10^{26}$~W of PSR~J1534+12 combined with its distance gives a figure-of-merit acceptable for Fermi/LAT \citep{2013ApJS..208...17A}. It should therefore be detected as a gamma-ray pulsar all the time. The fact that no pulsed gamma-ray emission has been reported by Fermi/LAT shows that the geometry derived by the rotating vector model might be inconsistent with the real geometry of this pulsar.  \cite{1996ApJ...470.1111A} found possible fits close to an orthogonal rotator, which contradicts the fact that a large off-pulse component is present, which would favour an aligned rotator. The geometric parameters used in these simulations probably do not correspond to the true geometry of the pulsar. The complex profile morphology related to a complex polarization angle evolution seems to lead to models that include multipolar components. Our recent work in progress demonstrates that an almost aligned rotator including a quadrupolar component could be misinterpreted as an orthogonal dipole rotator. 

\subsection{PSR~J1141-6545}

\begin{figure*}
 \centering
\begin{tabular}{cc}
 \includegraphics[width=0.45\textwidth]{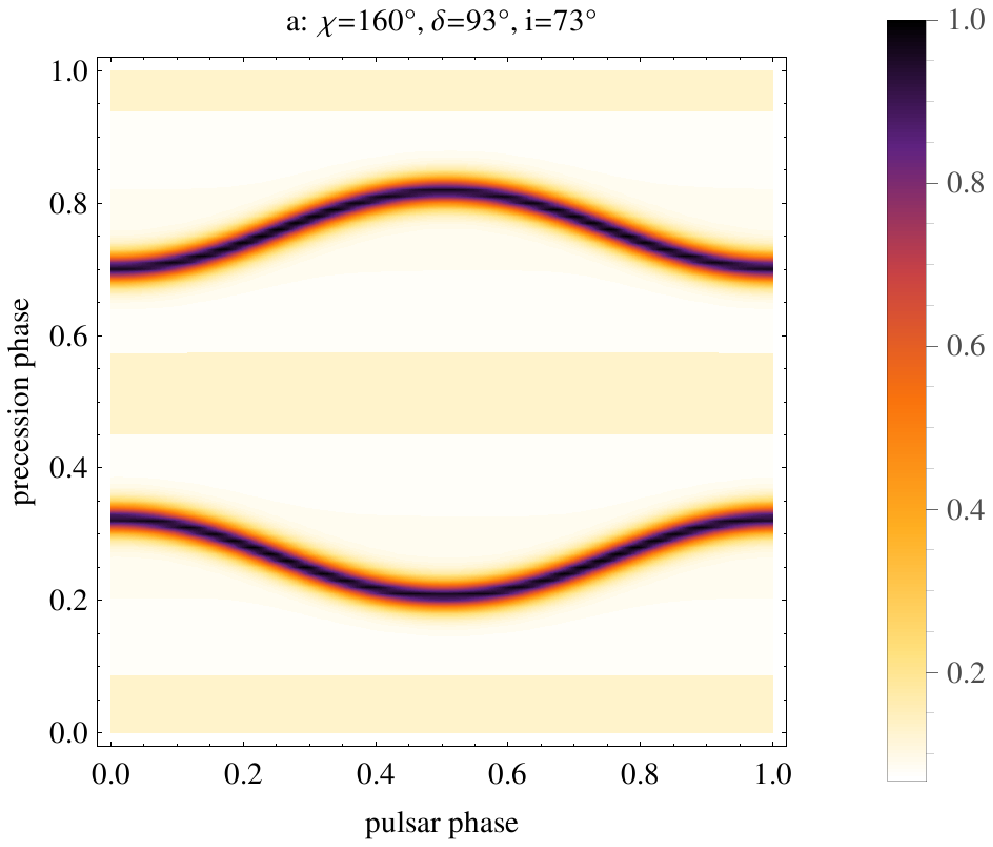} &
 \includegraphics[width=0.45\textwidth]{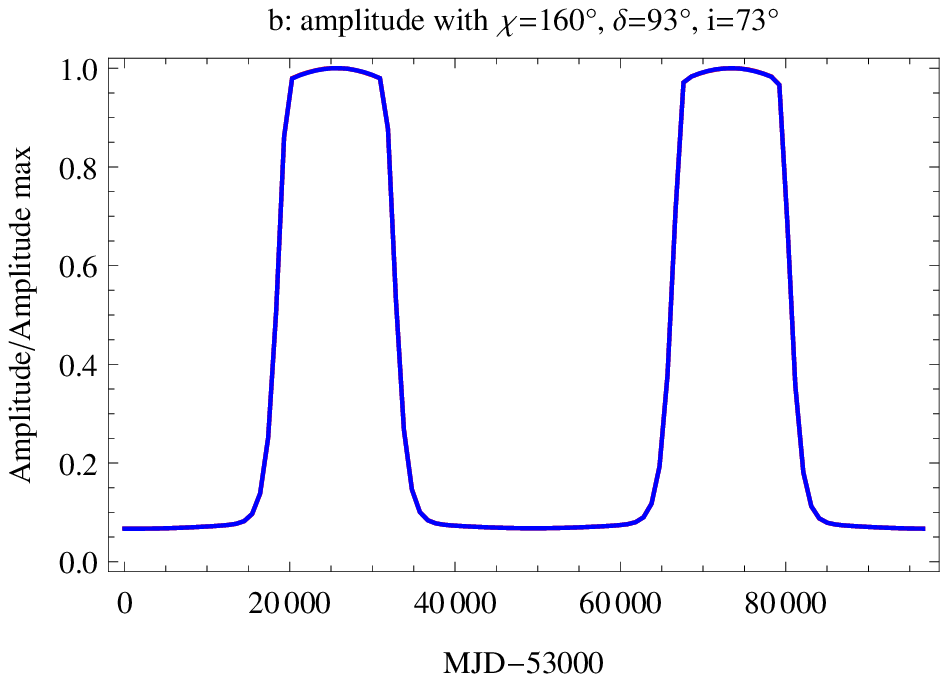} \\
 \includegraphics[width=0.45\textwidth]{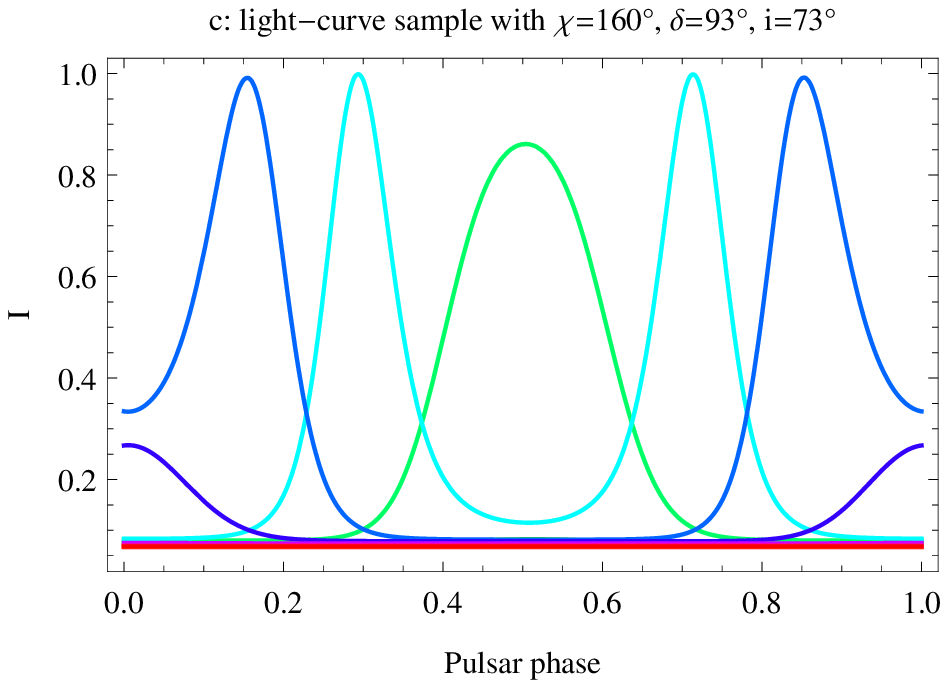} &
 \includegraphics[width=0.45\textwidth]{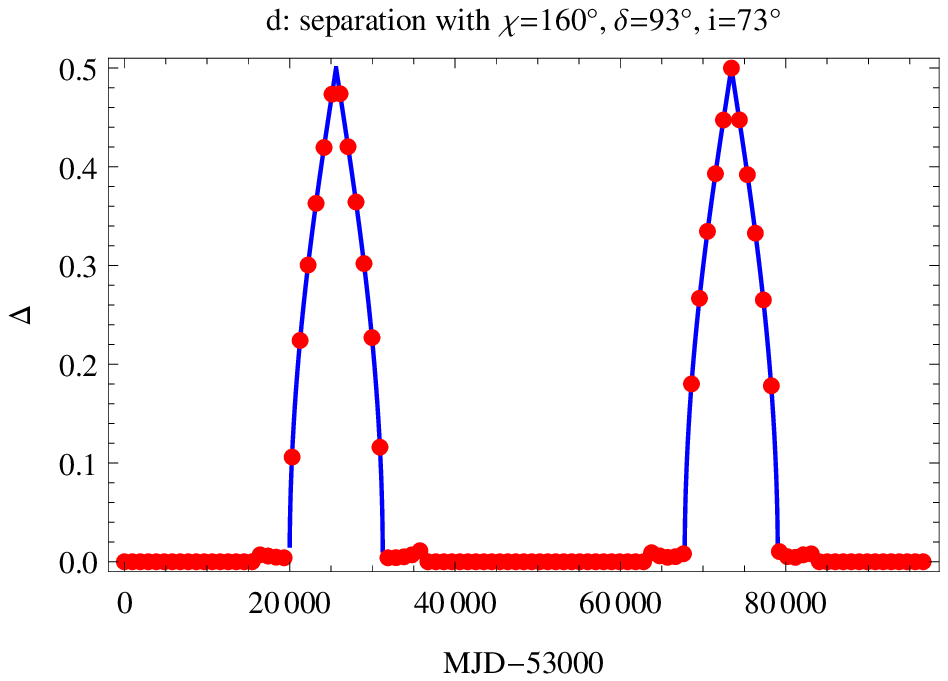}
\end{tabular}
 \caption{Detailed analysis of the evolution of the pulsed emission from PSR~J1141-6545. From left to right and top to bottom, we show the pulse profiles, the amplitude of the peak intensity, the width of the pulses, and the separation of the two peaks. Note the different x-label, pulsar phase for the light curves but precession phase for the other plots, here expressed in MJD.}
 \label{fig:PSRJ1141}
\end{figure*}

PSR~J1141-6545 is a binary system of a pulsar and a white dwarf in close and eccentric orbit \citep{2011MNRAS.412..580A}. The neutron star has a period of $P=394$~ms and a period derivative of $\dot{P}=4\times10^{-15}$. Its geodetic precession rate has been estimated by \citet{2005ApJ...624..906H}. \cite{2010ApJ...710.1694M} also used the rotating-vector model to fit the data. They derived $\chi=160^o$, $i=73^o$, and $\delta=93^o$. This high misalignment makes this system a good candidate for pulsed X-rays/gamma-rays emission. Therefore, we studied its pulse profile evolution in more detail with the values given by \cite{2010ApJ...710.1694M}, including the precession angle $\Phi_0$ at Modified Julian Day (MJD)~53000.0. Results are shown in figure~\ref{fig:PSRJ1141}. The amplitude of the pulsations remains weak until approximately MJD~68000, corresponding to the year around~2045, where it will significantly increase (panels~\ref{fig:PSRJ1141}a,b). If PSR~J1141-6545 is an X-ray/gamma-ray pulsar, it should then be detected starting with one pulse followed by a shift to a double-pulse structure. Its spin-down luminosity close to $3\times10^{26}$~W, as for the previous pulsar, lies on the lower limit of the Fermi/LAT rotation-powered pulsar detection. PSR~J1141-6545 should therefore be detected as a high-energy pulsar, at least marginally, in a few decades.

\subsection{Comparison with outer gap and two-pole caustic predictions}

The location of the site of high-energy photon production is one of the key questions in pulsar gamma-ray emission. We presented calculations in the wind-zone scenario at distances larger than the light-cylinder radius. The striped-wind model has been successfully applied to the pulsed optical polarization in the Crab pulsar \citep{2005ApJ...627L..37P}, the discovery of the Crab flares in the nebula or in the unshocked wind \citep{2013MNRAS.436L..20B}, the gamma-ray luminosity of Fermi/LAT pulsars \citep{2012MNRAS.424.2023P}, the sharpening of the pulse profile when moving to higher energies \citep{2011MNRAS.417..532P}, as seen for instance in the Vela pulsar \citep{Abdo2010} and a possible solution to the sigma problem in pulsar winds with strong magnetic dissipation at the termination shock \citep{2007A&A...473..683P}.

It will be interesting to see if future observations of this geodetic precession can exclude some high-energy emission models like the outer gap or the two-pole caustic model. To bring some decisive arguments to this discussion, we compare in this last section the striped-wind predictions with the popular outer gap and two-pole caustic expectations. We show that the different models disagree strongly, leading to predicted light-curves evolution according to geodetic precession that are incompatible between the models. This discrepancy will eventually lift the degeneracy between several scenarios and favour one of these models. Our two-pole caustic model follows \cite{2003ApJ...598.1201D}. Emission occurs along the last closed field line from the neutron star surface up to a substantial fraction of the light-cylinder radius~$\rlight$. More precisely, we took $R_{\rm max} = \rho_{\rm max}=0.95\,\rlight$; see the above mentioned work for a detailed discussion of the meaning of these quantities. For the outer gaps, a cavity in the magnetosphere can be excised in several ways, see for instance \cite{2000ApJ...537..964C}. We assumed a model with the least number of parameters: a reduced two-pole caustic model where emission starts when the magnetic field line crosses the null surface. This would represent a simplistic outer-gap model similar to that chosen by \cite{2003ApJ...598.1201D}. The retarded dipole field lines are given in spherical coordinates in the Appendix, see equation~(\ref{eq:DipoleRetarde}). The precise boundary of the outer gaps are not relevant in our study because we wish to investigate the evolution of the peak intensity and separation with the line-of-sight inclination. The details of the pulse profile such as its width are of no concern here because we are able to distinguish between the three scenarios by inspecting the peak separation and intensity with geodetic precession phase as we show below.

\begin{figure}
 \centering
\includegraphics[width=0.45\textwidth]{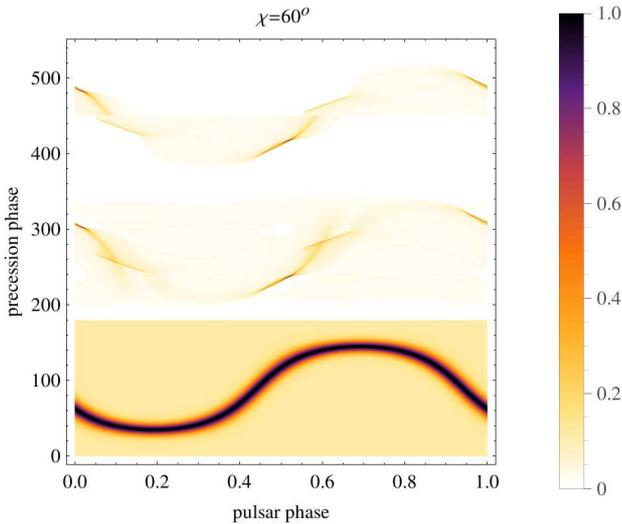}
 \caption{Phase plot diagram for our striped-wind model, in the lower panel for $\zeta\in[0^o,180^o]$, the two-pole caustic model in the middle panel for $\zeta\in[180^o,360^o]$, and the outer-gap model in the upper panel for $\zeta\in[360^o,540^o]$. The plots have been artificially shifted to place them in one graph. The pulsar obliquity is $\chi=60^o$.}
 \label{fig:Carte_TPC_OG}
\end{figure}

We first summarize the phase plot diagrams expected from the three models. A typical example is shown in figure~\ref{fig:Carte_TPC_OG} for the striped wind, the two-pole caustic, and the outer-gap model for an obliquity of the pulsar of $\chi=60^o$. The overall variation of the peak separation with respect to the inclination of the line of sight is very similar among the three models. However, the striped-wind model alone always yields symmetric pulses regardless of the inclination angle. The caustic nature of the other two phase plot diagrams leads to asymmetric pulses with peak intensity ratios evolving with $\zeta$. They also show several abrupt changes in their light-curves.

Then we briefly review the behaviour of the light curves for the two-pole caustic and the outer-gap models for each of the double neutron star system. Note that to facilitate the comparison among the phase plots from the striped-wind model with those of the outer-gap and two-pole caustics, we adjusted the pulsar rotational phase as well as the geodetic precession phase to obtain similar diagrams. There is indeed a freedom in the origin of the two phases because both motions are periodic, although obviously on very different time scales.

\begin{figure*}
 \centering
\begin{tabular}{cc}
\includegraphics[width=0.45\textwidth]{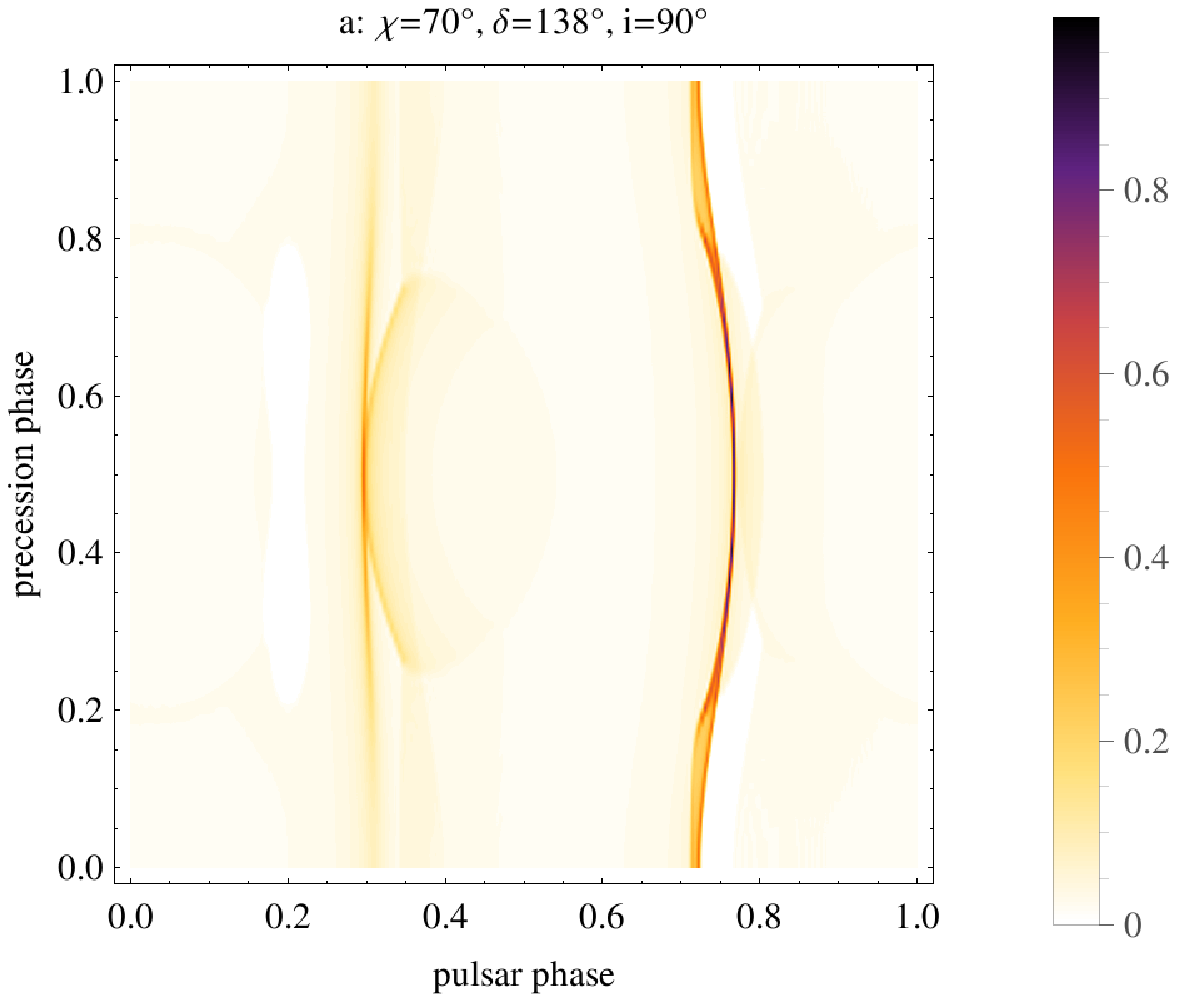} &
\includegraphics[width=0.45\textwidth]{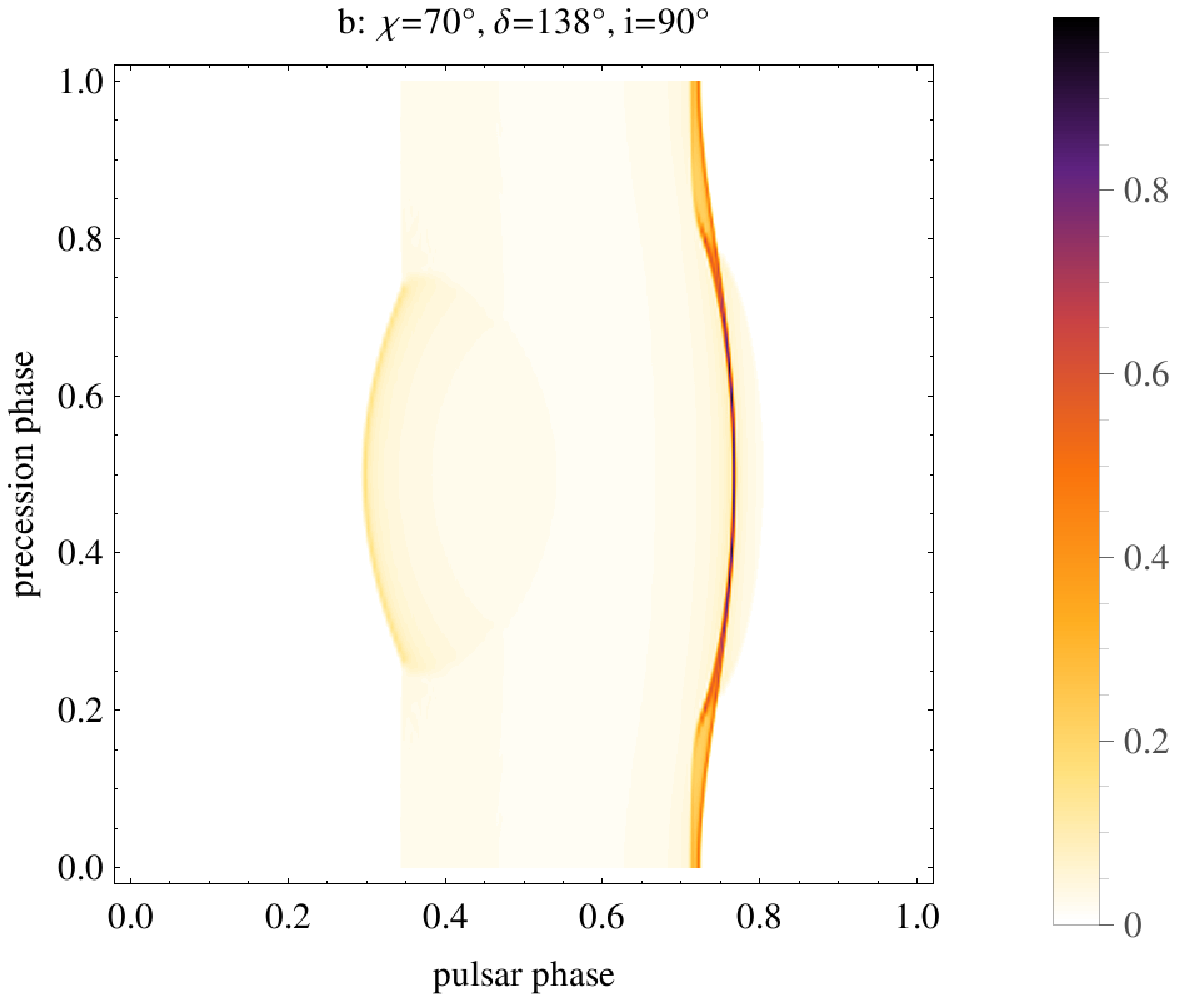}
\end{tabular}
 \caption{Evolution of the pulse profiles with respect to the precession phase~$\varphi$ for PSR~J0737-3039B. Predictions for the two-pole caustics at the left, and for the outer-gap model at the right.}
 \label{fig:Precession_c70_TPC_OG}
\end{figure*}

\begin{figure}
 \centering
\includegraphics[width=0.45\textwidth]{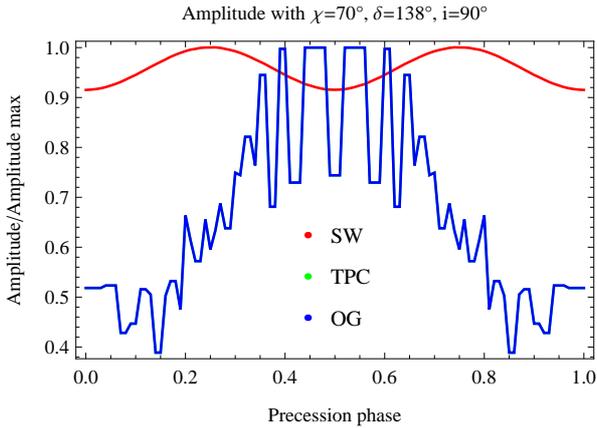}
 \caption{Evolution of the peak intensity with respect to the precession phase~$\varphi$ for PSR~J0737-3039B. Predictions for the striped-wind, the two-pole caustic, and outer-gap model are shown. TPC and OG overlap.}
 \label{fig:SW_TPC_OG_c70}
\end{figure}

First, for the double pulsar PSR~J0737-3039, the light-curve evolution of pulsar~B is shown in figure~\ref{fig:Precession_c70_TPC_OG} for the two alternatives, two-pole caustics on the left, and outer-gap model on the right. The two-pole caustic light-curve evolution resembles the striped-wind map. Two pulses are seen for most of the precession period. While the first peak remains approximately at the same rotational phase of $\varphi\approx0.3$, the second peak shifts to later phase, with changing peak intensity, from $\varphi\approx0.7$ to $\varphi\approx0.8$, as in figure~\ref{fig:PSRJ0737}. Thus the discrepancy in the phase shift of the first peak which is about~10\% for the striped-wind model, could be reported by current instruments if observed. The outer-gap map is similar to the two-pole caustic, except that part of the emission has been suppressed because of the null surface constraint. In that case, the peak separation increases and decreases again, as for the striped wind. But the pronounced asymmetry in the light curve should also be detected by current telescopes. 
In figure~\ref{fig:SW_TPC_OG_c70}, we report the peak intensity evolution with respect to the geodetic precession phase for all three models: striped wind (SW), two-pole caustic (TPC) and outer gap (OG). TPC and OG curves overlap, they are indiscernible. The intensity fluctuations for the SW are about~10\% while they are about~50\% for the TPC and OG. These higher variations for the magnetospheric emission models could easily be detected by any instrument.

\begin{figure*}
 \centering
\begin{tabular}{cc}
\includegraphics[width=0.45\textwidth]{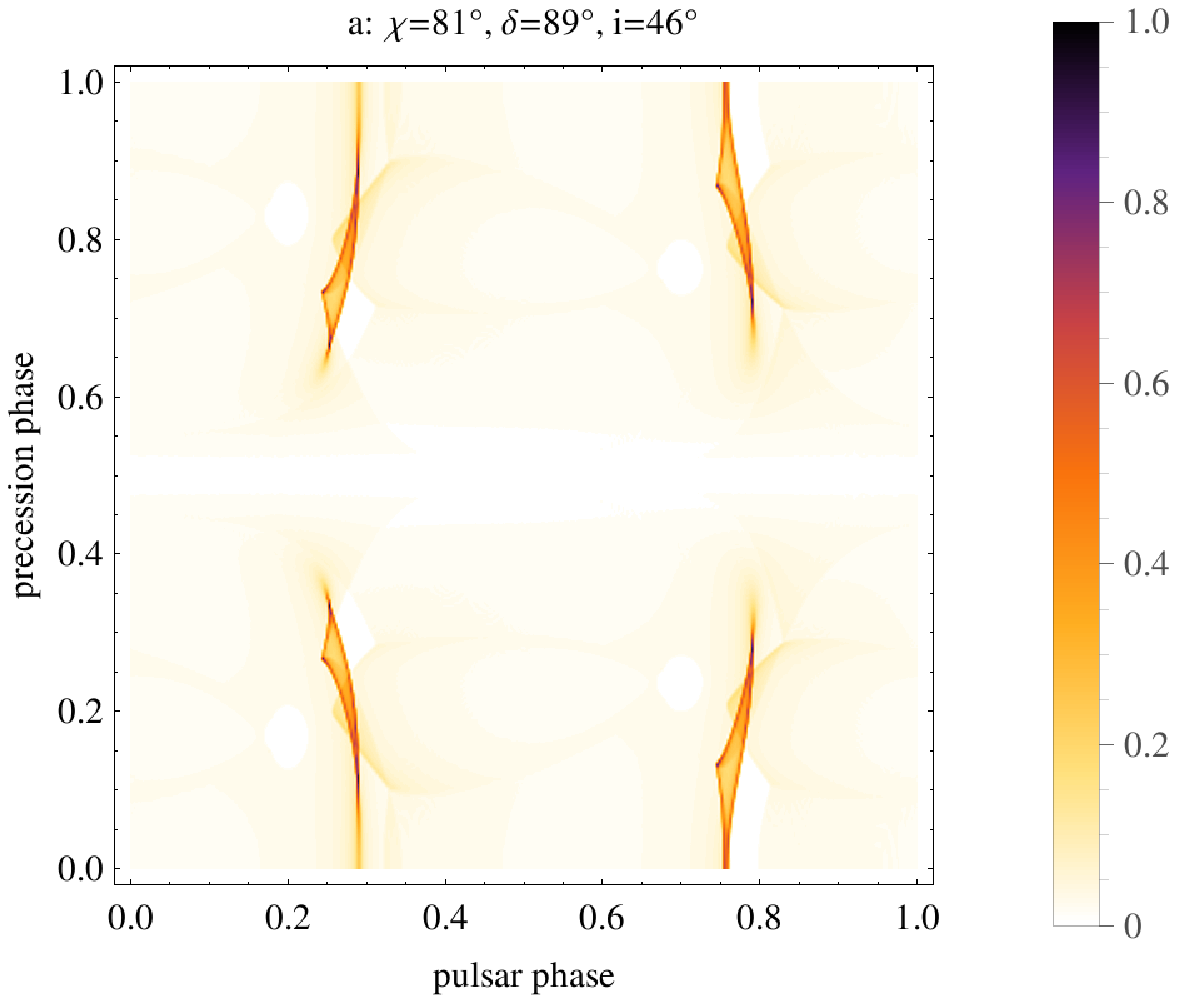} &
\includegraphics[width=0.45\textwidth]{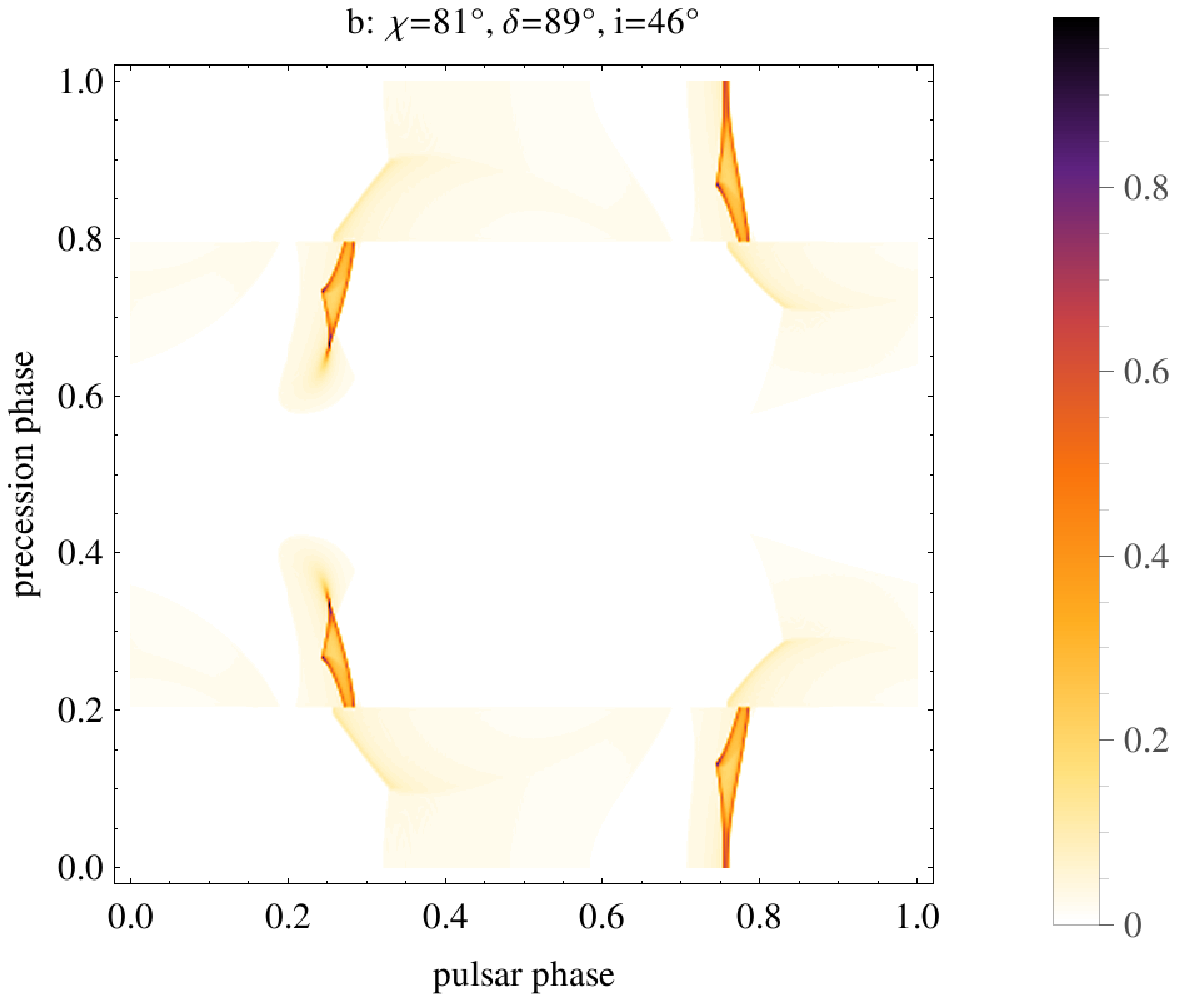}
\end{tabular}
 \caption{Evolution of the pulse profiles with respect to the precession phase~$\varphi$ for PSR~J1906+0746. Predictions for the two-pole caustics at the left, and for the outer-gap model at the right.}
 \label{fig:Precession_c81_TPC_OG}
\end{figure*}

Now we consider PSR~J1906+0746, whose light-curve evolution is shown in figure~\ref{fig:Precession_c81_TPC_OG}. Its SW light-curve map resembles the previous one. Two peaks are always present, but their separation increases during geodetic precession, as in figure~\ref{fig:PSRJ1906}. The TPC shows a similar trend with two peaks. But there is an essential difference between the two. TPC predicts a long geodetic precession phase where the pulsation disappears between a phase in which two pulses appear. In the OG expectations, only one pulse is produced, regardless of the precession phase. This single pulse disappears at the same precession phase as in the TPC model.

\begin{figure}
 \centering
\includegraphics[width=0.45\textwidth]{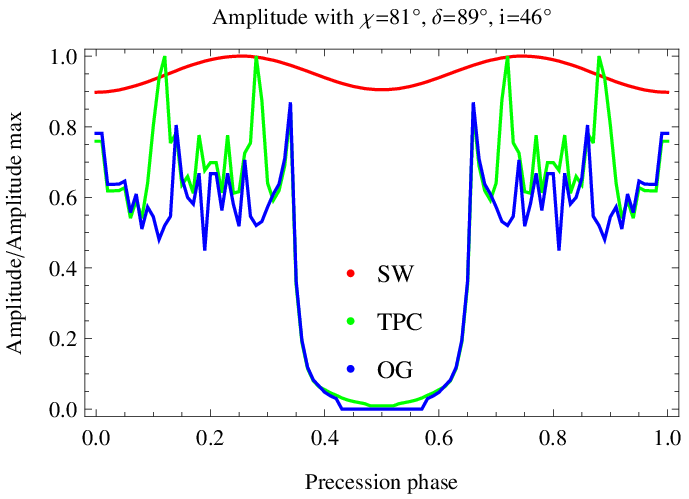}
 \caption{Evolution of the peak intensity with respect to the precession phase~$\varphi$ for PSR~J1906+0746. Predictions for the two-pole caustics at the left, and for the outer-gap model at the right. Note the significant extinction around $\varphi=0.5$ for the TPC and OG.}
 \label{fig:SW_TPC_OG_c81}
\end{figure}

The peak intensity in figure~\ref{fig:SW_TPC_OG_c81} clearly shows the distinction emphasized in the phase plots between SW and TPC/OG. A strong decrease in the pulsed intensity from that pulsar would be a clear signature of magnetospheric emission models and would rule out the SW.

\begin{figure*}
 \centering
\begin{tabular}{cc}
\includegraphics[width=0.45\textwidth]{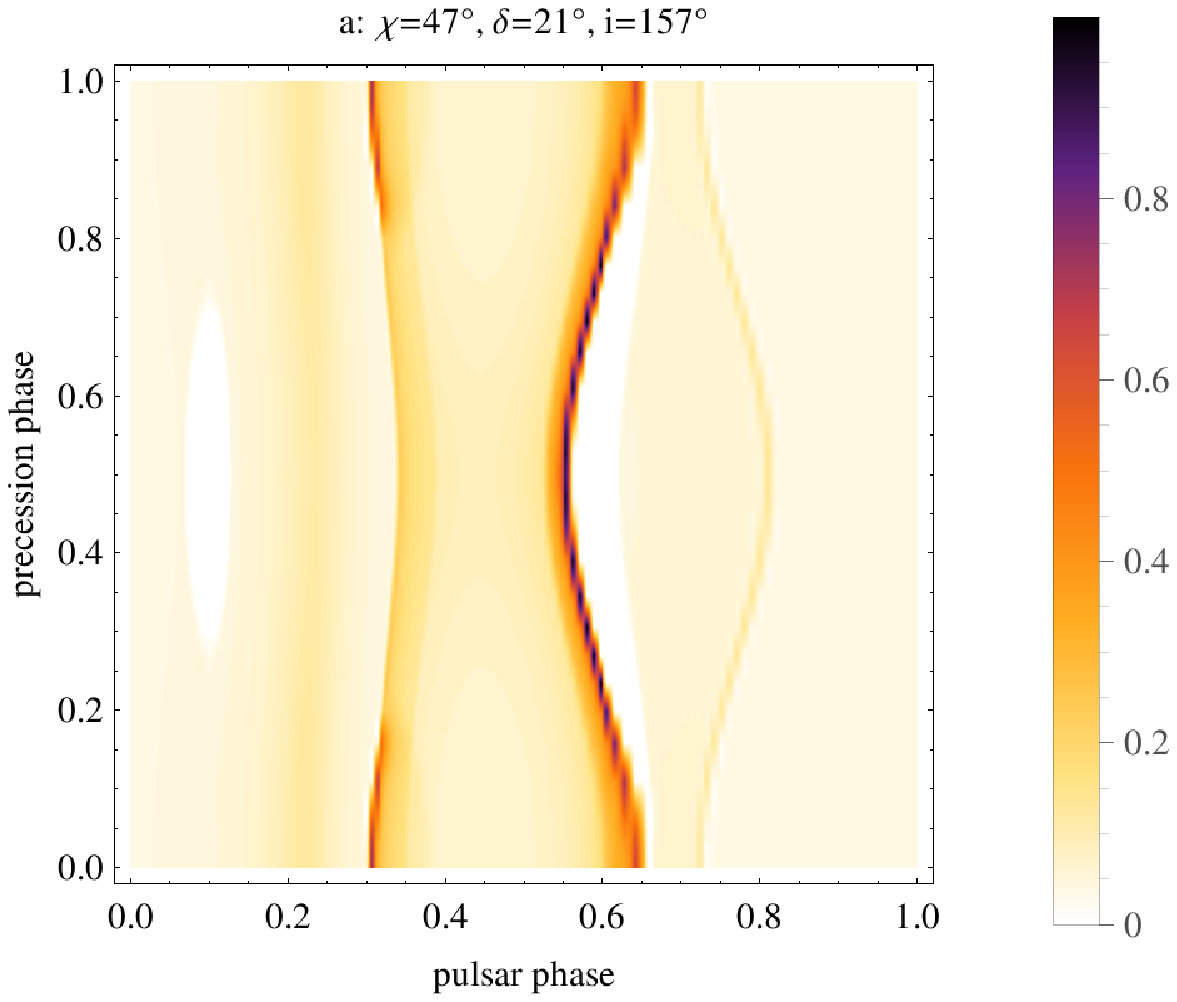} &
\includegraphics[width=0.45\textwidth]{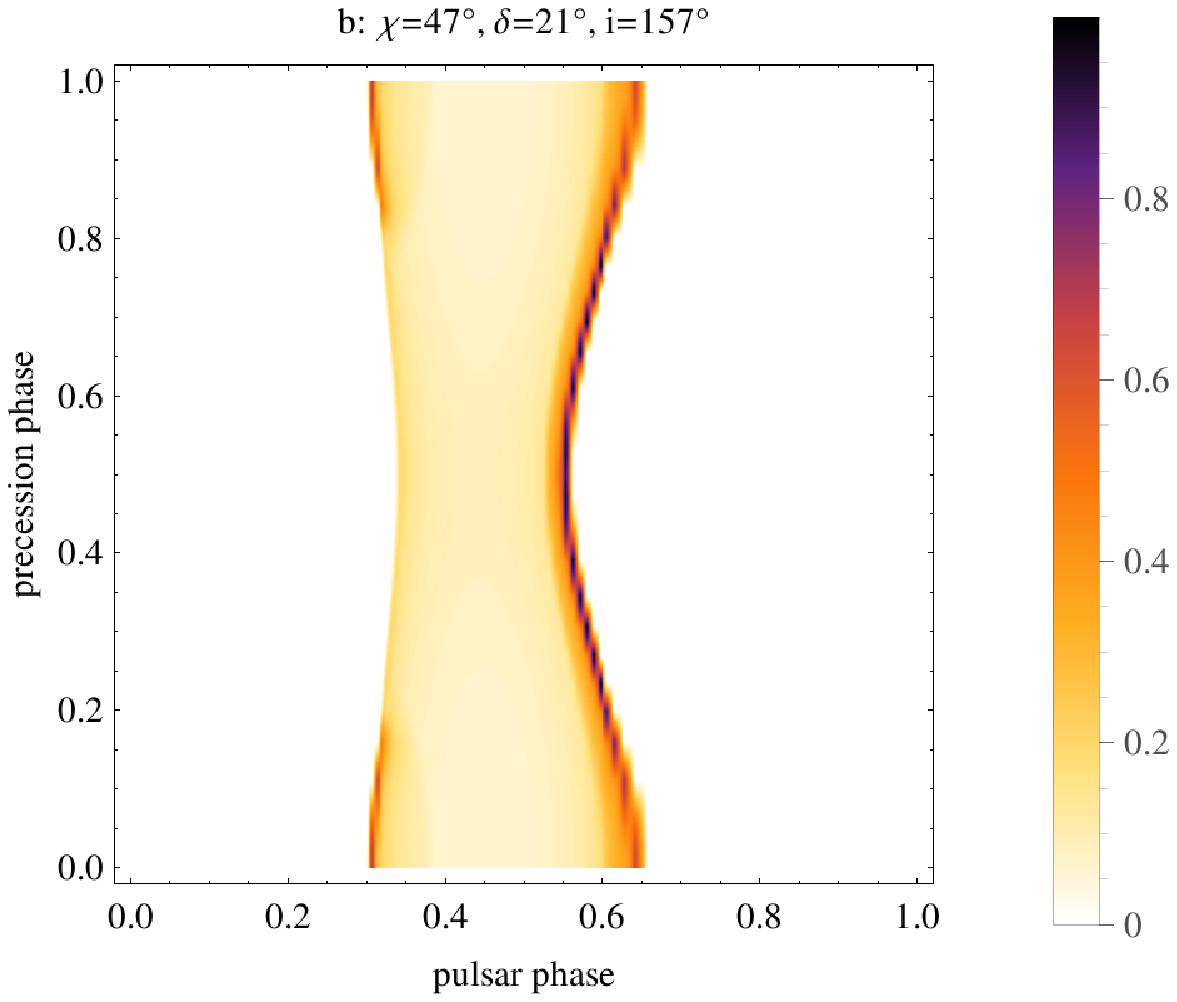}
\end{tabular}
 \caption{Evolution of the pulse profiles with respect to the precession phase~$\varphi$ for PSR~B1913+16. Predictions for the two-pole caustics at the left, and for the outer-gap model at the right.}
 \label{fig:Precession_c47_TPC_OG}
\end{figure*}

\begin{figure}
 \centering
\includegraphics[width=0.45\textwidth]{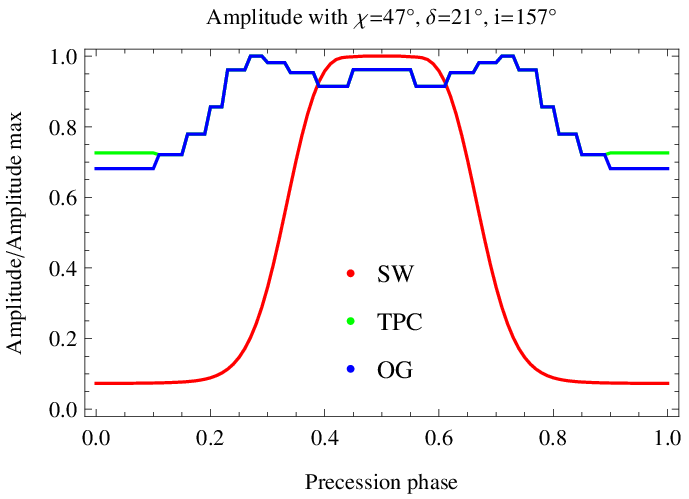}
 \caption{Evolution of the peak intensity with respect to the precession phase~$\varphi$ for PSR~B1913+16. Predictions for the two-pole caustics at the left, and for the outer-gap model at the right.}
 \label{fig:SW_TPC_OG_c47}
\end{figure}

We proceed with PSR~B1913+16, whose light-curve evolution is plotted in  figure~\ref{fig:Precession_c47_TPC_OG}. While the SW only shows one long pulse preceded by a long quiet phase, the TPC and OG always produce two pulses at all precession phases. Thus here again, it would be easy to distinguish the emission process from the different models. In the SW, the single pulse does not shift in phase. In the TPC and OG, the double-peak structure evolves in phase with a decrease from $\varphi\approx 0.65$ to $\varphi\approx 0.55$ for the brightest pulse and strong intensity fluctuations for the faintest pulse.
The peak intensity evolution in figure~\ref{fig:SW_TPC_OG_c47} makes the criteria used to distinguish the models even clearer. On one hand, the SW produces a 90\% variation in the emitted flux with total disappearance of pulsation and only one DC component. On the other hand, the TPC and OG only produce a 30\% variation in flux and pulsations throughout the geodetic precession.

\begin{figure*}
 \centering
\begin{tabular}{cc}
\includegraphics[width=0.45\textwidth]{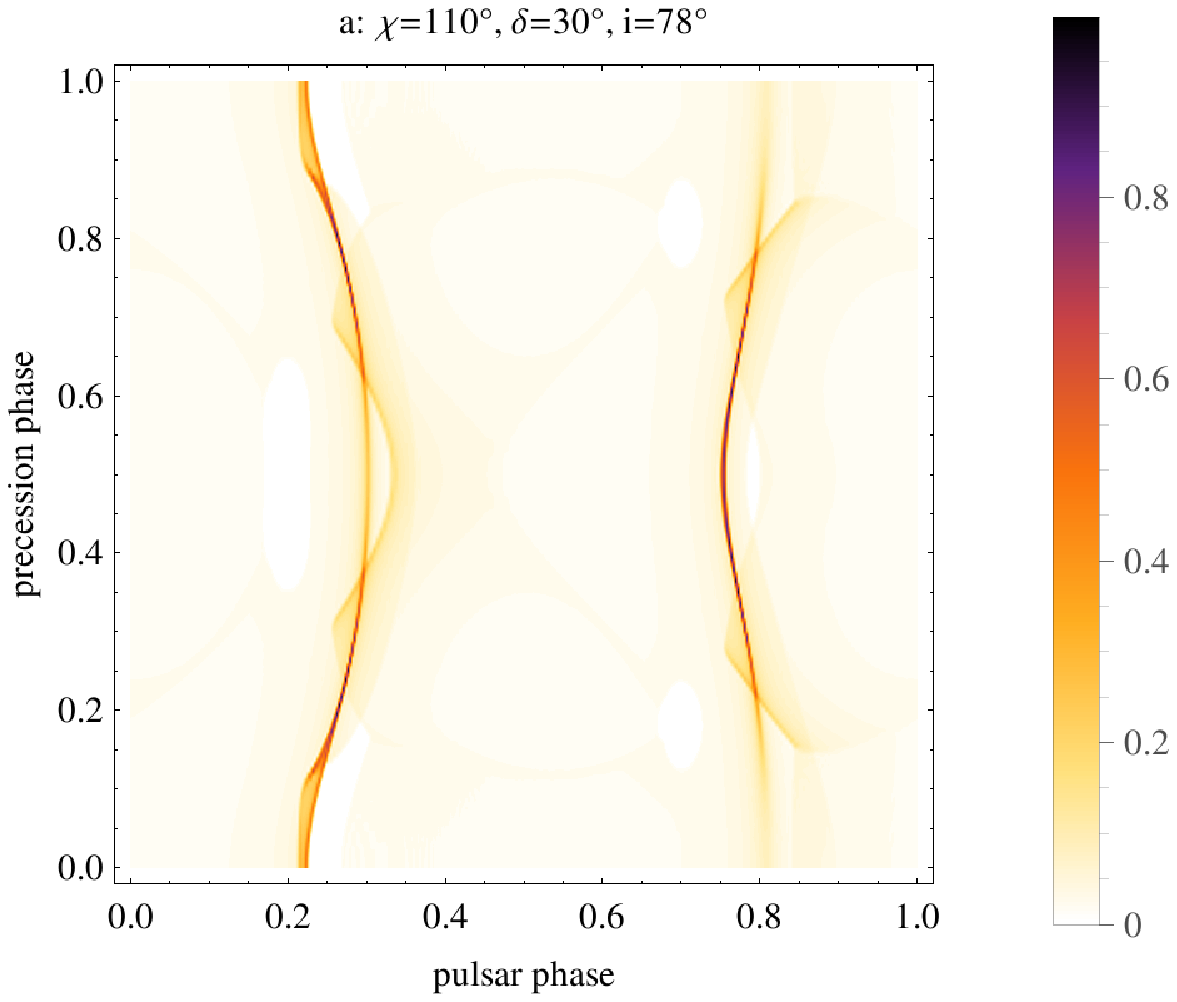} &
\includegraphics[width=0.45\textwidth]{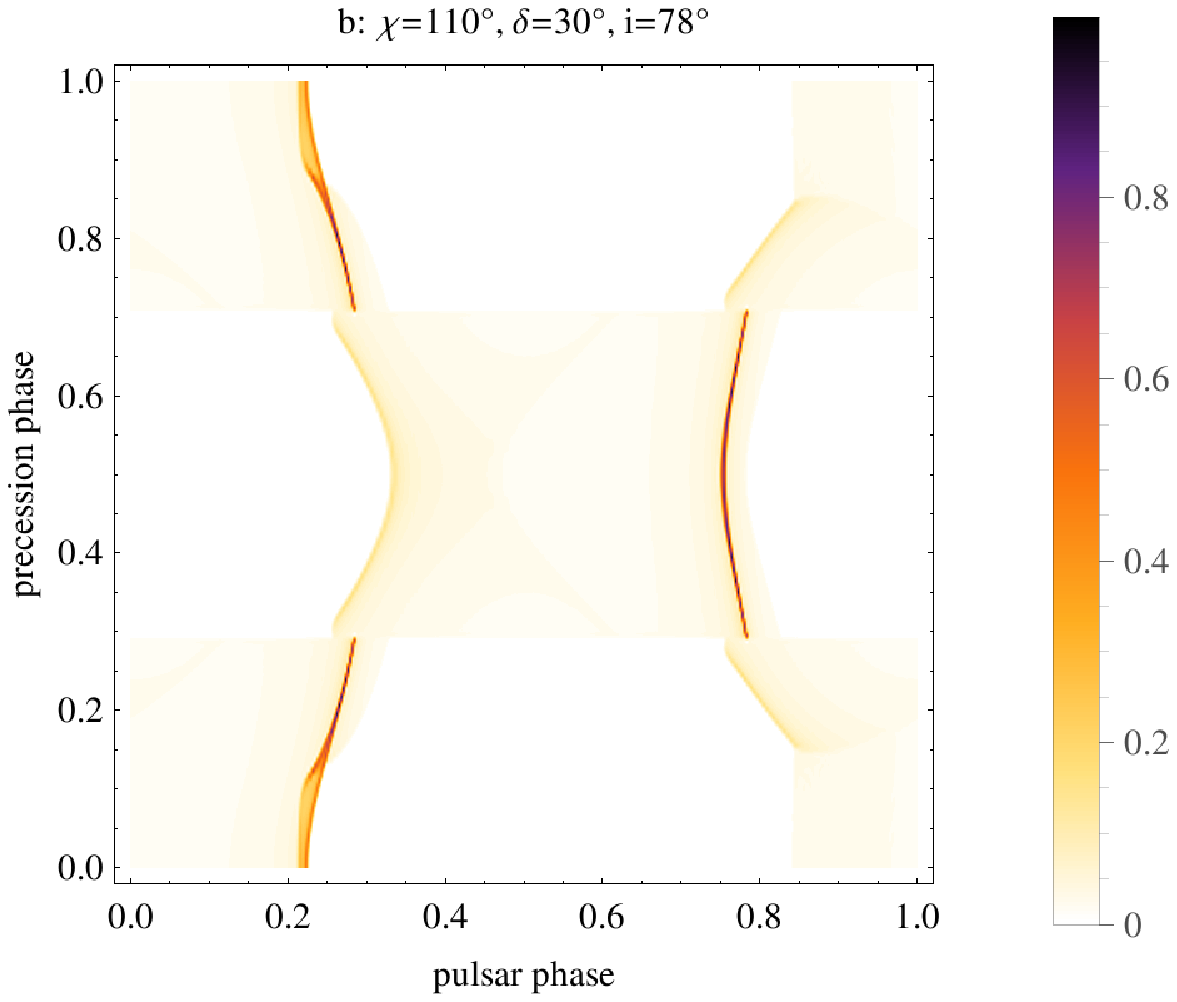}
\end{tabular}
 \caption{Evolution of the pulse profiles with respect to the precession phase~$\varphi$ for PSR~B1534+12. Predictions for the two-pole caustics at the left, and for the outer-gap model at the right.}
 \label{fig:Precession_c110_TPC_OG}
\end{figure*}

\begin{figure}
 \centering
\includegraphics[width=0.45\textwidth]{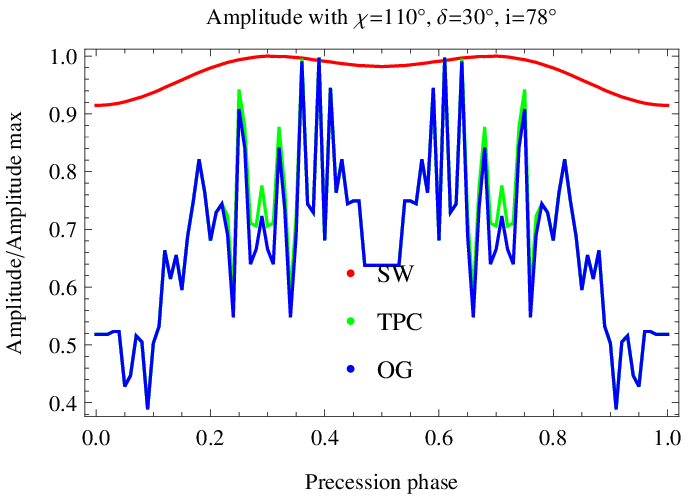}
 \caption{Evolution of the peak intensity with respect to the precession phase~$\varphi$ for PSR~B1534+12. Predictions for the two-pole caustics at the left, and for the outer-gap model at the right.}
 \label{fig:SW_TPC_OG_c110}
\end{figure}

For PSR~B1534+12, figure~\ref{fig:Precession_c110_TPC_OG}, the cut between the models is less clear. The SW and the TPC diverge in their peak intensity, but not in the pulse profile separation. They decrease in the same manner for both peaks. For the OG, the single peak during the full geodetic precession phase would be a strong indicator for magnetospheric emission in cavities located close to the light-cylinder.
However, the peak intensity evolution will not be as good an indicator to distinguish among the models, as concluded from figure~\ref{fig:SW_TPC_OG_c110}.

\begin{figure*}
 \centering
\begin{tabular}{cc}
\includegraphics[width=0.45\textwidth]{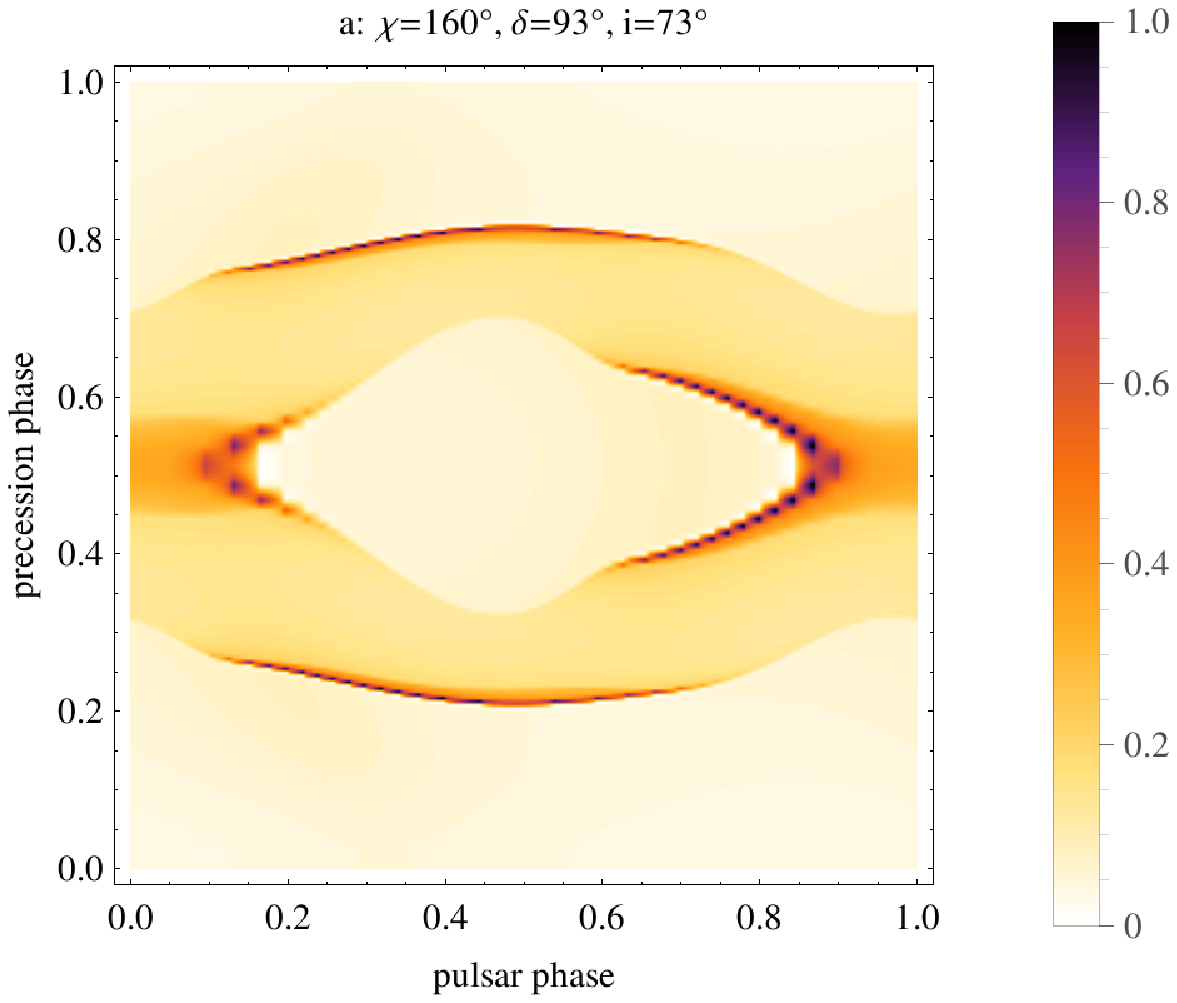} &
\includegraphics[width=0.45\textwidth]{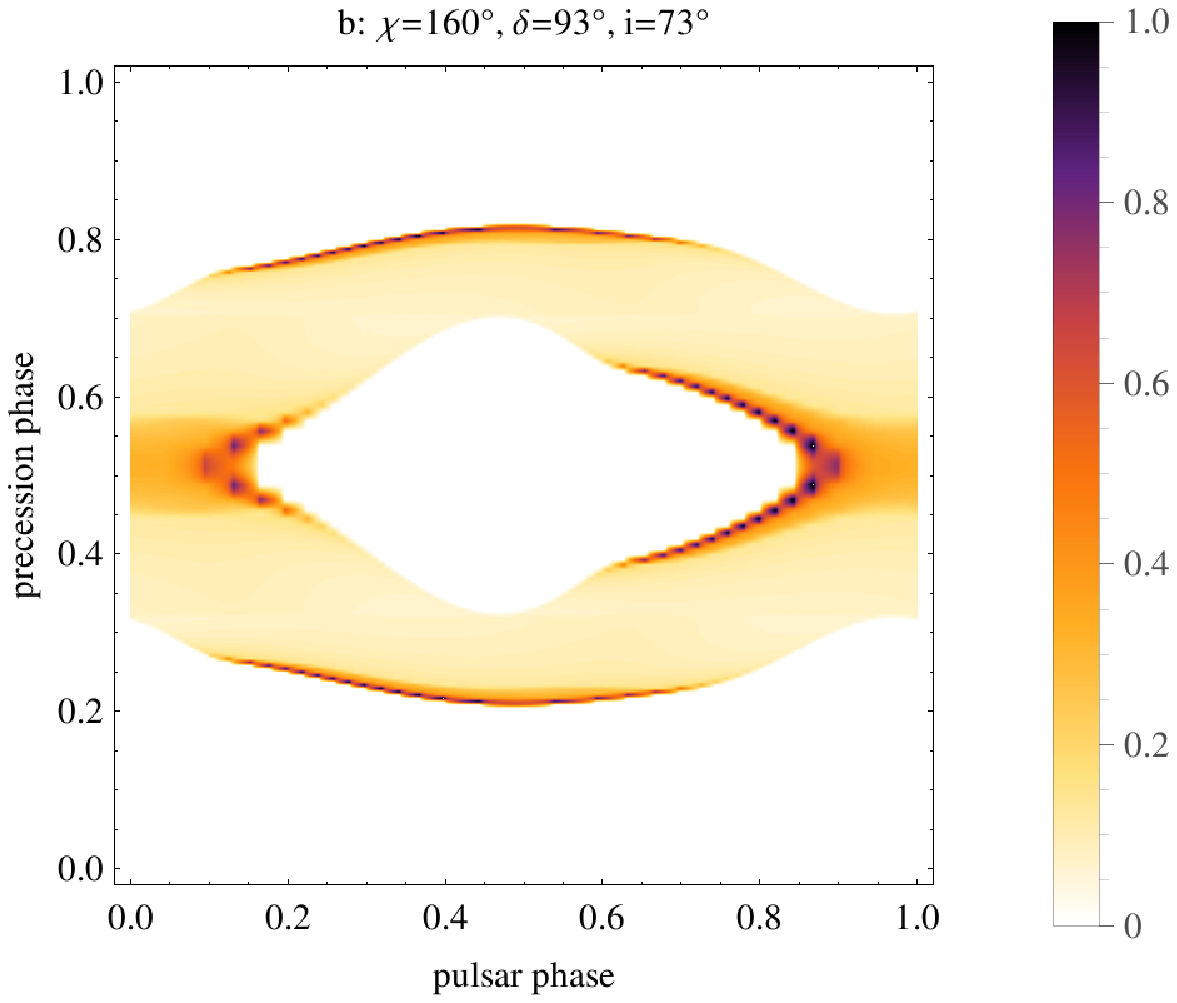}
\end{tabular}
 \caption{Evolution of the pulse profiles with respect to the precession phase~$\varphi$ for PSR~J1141-6545. Predictions for the two-pole caustics at the left, and for the outer-gap model at the right.}
 \label{fig:Precession_c160_TPC_OG}
\end{figure*}

\begin{figure}
 \centering
\includegraphics[width=0.45\textwidth]{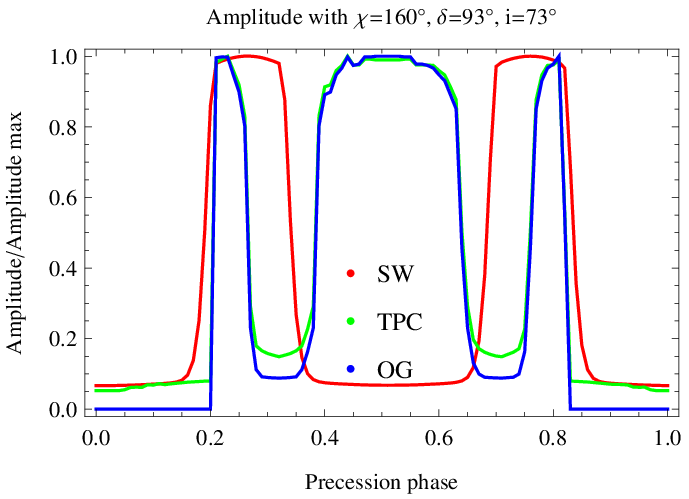}
 \caption{Evolution of the peak intensity with respect to the precession phase~$\varphi$ for PSR~J1141-6545. Predictions for the two-pole caustics at the left, and for the outer-gap model at the right.}
 \label{fig:SW_TPC_OG_c160}
\end{figure}

Finally, we conclude with PSR~J1141-6545, whose light-curve evolution is given in figure~\ref{fig:Precession_c160_TPC_OG}. After a quiescent phase, a single pulse appears and separates to form a double-peak profile in all models. So the SW, TPC, and OG models look very similar at first sight. But around geodetic precession phase $\varphi\approx0.5$, the TPC and OG produce significant pulsed emission, which is not the case for the SW. 
This fact is clearly seen in figure~\ref{fig:SW_TPC_OG_c160}. The SW passes through two strong emission phases at $\varphi\in[0.15,0.35]$ and $\varphi\in[0.65,0.85]$, whereas the TPC and OG models possess three distinct emission phases, one centred around 0.5, which Moreover is absent in the SW, and two around $\varphi\in[0.2,0.3]$ and $\varphi\in[0.75,0.85]$.
This system is most probably the best candidate for probing pulsar high-energy emission models because of its high precession rate and unambiguous signature discrepancy between the models.

In concluding, we remark that if the phase plots cannot be detected to sufficiently high precision to extract light curves with high confidence, at least the fluctuations in gamma-ray fluxes integrated over a significant fraction of the geodetic precession period should be able to lift the degeneracy among competing models in some double neutron star systems in a near future (by near we mean within a researcher career which today is about 45~years!).

\section{CONCLUSION}
\label{sec:Conclusion}

We showed that geodetic precession imprints a clear signature on the evolution of high-energy pulse profiles in the striped-wind scenario. Although the geodetic precession rate is predicted from Eq.~(\ref{eq:Precession}) to very high accuracy through precise radio-timing observations, the inferred pulsar geometry from the radio pulse-profile changes remains uncertain. As yet, there is no clear confirmation of a similar trend in X-ray or gamma-ray pulse profiles. This non-detection is maybe intrinsic to the pulsar emission geometry, which could differ between radio and high-energy photons or simply because of some limitations in the Fermi/LAT construction of the light curves (limited signal-to-noise ratio, or too short observing baselines). Nevertheless, we expected that these phenomena will be visible in the near future for some double neutron stars. Such detections will help us constrain the high-energy pulsed emission models for pulsars, a long-standing question of pulsar physics, but still largely unsolved. PSR~J1906+0746 and PSR~J1141-6545 are the best candidates to look for this geodetic spin precession in gamma-ray energies and are accessible with current telescopes. Our predictions indicate that within a few decades this emission will be detected with existing technology. Our conclusions crucially depend on the parameters used for the magnetic obliquity and the inclination of the line of sight. Definitive conclusions can be drawn only after these parameters have been fixed without doubt, a task still to be accomplished.

Predictions from other competing models for this pulsed emission were performed. This study may help in modelling future decade-long observations of X-ray and gamma-ray pulsars. These measurements will definitely rule out some models. We encourage observers to look for such signatures in their data. To clearly point out the discrepancies between the two-pole caustic, the outer-gap, and the wind scenarios, we showed the evolution of the pulse profiles and highest peak intensity with respect to the geodetic precession phase. In some double neutron star systems, the predictions diverge deeply. Moreover, a consistency check between the geometry determined by radio observations and high-energy emission will provide another severe criterion to distinguish between models and their related magnetic topology.

Finally, we emphasize that all the work presented here relies on fits of the radio pulse profile and polarization with the rotating-vector model. Although it yields good results for some pulsars, many outliers exhibit a more complex pulse structure and polarization evolution. To obtain reliable fits, magnetic multipolar components could play a central role. We currently work on this problem to determine to which extent it could modify the classical dipolar geometry and what its implications are for pulsar magnetospheric physics.

\section*{Acknowledgements}

This work has been supported by the French National Research Agency (ANR) through the grant No. ANR-13-JS05-0003-01 (project EMPERE) and the Programme National Hautes \'Energies (PNHE). It also benefited from the computational facilities available at Equip@Meso (Universit\'e de Strasbourg). I am grateful to Lucas Guillemot for stimulating discussions.

\appendix
 
\section{Retarded point dipole}

In the slot-gap and outer-gap models, the precise structure of the magnetic field is essential for understanding the pulse profile evolution with the pulsar rotational phase. Indeed, their high-energy emission relies on curvature radiation along some open magnetic field lines. Aberration and retardation effects are central to produce caustics, as explained by \cite{1983MNRAS.202..495M}. To complete our picture of the outer-gap and two-pole caustic model used in this paper, we briefly recall the components of the retarded magnetic dipole radiation. We specify it to the limit of vanishing stellar radius by setting $R\to0$ in the Deutsch solution (but assuming a constant magnetic moment $\mu=B\,R^3$ in this limit) given for instance in full length in \cite{2012MNRAS.424..605P}. We obtain the following expressions in spherical coordinates
\begin{subequations}
\label{eq:DipoleRetarde}
 \begin{align}
  B_r & = \frac{2\,B\,R^3}{r^3} \, \left[ \cos\chi\,\cos\vartheta + \right. \\
  & \left. \sin\chi\,\sin \vartheta \, ( \cos(k\,r-\Omega\,t+\varphi) + k\,r\,\sin(k\,r-\Omega\,t+\varphi) ) \right] \nonumber \\
  B_\vartheta & = \frac{B\,R^3}{r^3} \, \left[ \cos\chi\,\sin\vartheta + \sin\chi\,\cos \vartheta \, \right. \\
  & \left. \left\{ (k^2\,r^2-1) \, \cos(k\,r-\Omega\,t+\varphi) - k\,r\,\sin(k\,r-\Omega\,t+\varphi) \right\} \right] \nonumber \\
  B_\varphi & = - \frac{B\,R^3}{r^3} \, \sin\chi\, \left[ k\,r\, \cos(k\,r-\Omega\,t+\varphi) + \right. \\
  & \left. (k^2\,r^2-1)\,\sin(k\,r-\Omega\,t+\varphi) \right] , \nonumber
 \end{align}
\end{subequations}
where $k=\Omega/c=1/\rlight$ is the wavenumber. These components are used to compute the magnetic field lines from which emission is supposed to come. Compared with a static dipole, corrections are provided by the $k\,r$~terms, which are of the order of unity close to the light cylinder.

\end{document}